\documentclass[aps,prl,reprint,groupedaddress,nofootinbib]{revtex4-1}

\usepackage[bottom]{footmisc}
\usepackage{amsmath}
\usepackage{slashed}
\usepackage{amssymb}
\usepackage{graphicx}
\usepackage{xcolor}

\newcommand{\Lagr}{\mathcal{L}}
\def\bea{\begin{eqnarray}}
\def\eea{\end{eqnarray}}
\def\hyp{\mathsf{y}}

\newcommand{\cT}{\mathcal T}
\newcommand{\cS}{\mathcal S}

\begin{document}

\title{Equations of motion, symmetry currents and EFT below the electroweak scale}
\author{Andreas Helset and Michael Trott}

\affiliation{Niels Bohr International Academy and Discovery Center,
Niels Bohr Institute,
University of Copenhagen, Blegdamsvej 17, DK-2100 Copenhagen, Denmark}


\date{\today}

\begin{abstract}
The low-energy effective field theory is constructed by integrating out Standard Model states with masses proximate to the electroweak scale.
We report the equations of motion for this theory, including corrections due to higher dimensional operators up to mass dimension six.
We construct the corresponding symmetry currents,
and discuss how the $\rm SU(2)_L \times U(1)_{\hyp}$ symmetry, and global symmetries, are manifested when
Standard Model states are integrated out. 
Including contributions from higher dimensional operators to the equations of motion
modifies the interpretation of conserved currents.
We discuss the corrections to the electromagnetic current as an example, showing how modifications
to the equation of motion, and corresponding surface terms, have a direct interpretation in terms of multipole charge distributions
that act to source gauge fields. 
\end{abstract}


\maketitle

{ \bf Introduction.}
Assuming physics beyond the Standard Model (SM) at scales
$\Lambda > \bar{v}_T = \sqrt{2 \, \langle H^\dagger H\rangle}$,
the embedding of the discovered ``Higgs-like" scalar into an $\rm SU(2)_L$ scalar doublet ($H$), and
the absence of hidden states with couplings to the SM and masses $ \lesssim  \bar{v}_T$,
the SM can be extended into the Standard Model Effective Field Theory (SMEFT).
Current LHC results are consistent with interpreting data in this framework, where an infinite tower of higher dimensional operators
is added to the SM.
The lack of any direct discovery of new physics resonances indicating beyond the SM states with masses $\sim \bar{v}_T$ also supports the assumption that
$\bar{v}_T/\Lambda < 1$. As a result, the SMEFT expansion in terms of local contact operators is a useful and predictive formalism to employ studying measurements
with characteristic scales $\sim \bar{v}_T$.

The SMEFT has the same field content as the SM, and reduces to the later by taking $\Lambda \rightarrow \infty$.
As the SM is falsified due to the evidence of neutrino masses from neutrino oscillations,
we assume that neutrino masses are generated by the dimension five SMEFT operator.

The LHC is providing large amounts of data measured around the scale $\bar{v}_T$ to search indirectly for physics beyond the SM.
These efforts are important to combine with experimental measurements at scales $\ll \bar{v}_T$,
where the Low-Energy Effective Field Theory (LEFT) is the appropriate EFT description.\footnote{The notation $\bar{v}_T$ indicates that this expectation value includes the effects
of possible higher dimensional operators.}
The LEFT is built out of the field content of the SM,  but as
the Higgs, $W^\pm$, $Z$, and top have masses $m_{W,Z,h,t} \sim \bar{v}_T$, these states are integrated out in sequence.
The gauged and linearly realized symmetries of the LEFT are $\rm U(1)_\textrm{em}$ and $\rm SU(3)_c$.
To perform EFT studies that combine data sets at scales $\sim \bar{v}_T$ and  $\ll \bar{v}_T$, one matches the SMEFT onto the LEFT, and uses renormalization group evolution to run between the different scales. For recent results to this end,
see Refs.~\cite{Jenkins:2017jig,Jenkins:2017dyc}.

When considering matching onto the LEFT at sub-leading order, it is usually necessary to take into account corrections to the equations of motion (EOM) that occur due to the local contact operators present in this theory. 
In  Ref.~\cite{Barzinji:2018xvu}, such corrections for the SMEFT were determined. In this paper, we determine these corrections for the LEFT up to  operators of mass dimension six.

The pattern of local operator corrections to the EOM encodes a (non-manifest) $\rm SU(2)_L \times U(1)_{\hyp}$ symmetry, when this
symmetry is assumed to be present in the UV completion of the LEFT. In this paper, we also construct the corresponding symmetry currents
and explain the way that the SM gauge symmetries, and global symmetries such as lepton number, are encoded in the LEFT.

Modifying the equations of motion of SM fields by higher dimensional operators
challenges the standard interpretation of conserved currents which is appropriate for, and limited to, renormalizable theories.
The generalized currents encode symmetry constraints that still constrain an EFT.
We also discuss how higher dimensional operator
corrections to the equation of motion have a direct interpretation in terms of multipole charge distributions
that act to source the corresponding gauge fields. We use the electromagnetic current as an example of this phenomena, and redefine 
the source in Gauss's law.

{\bf Effective field theory taxonomy.}
This paper is concerned with the connection between three effective theories: the Standard Model, the SMEFT and the LEFT.
Our SM notation is defined in Ref.~\cite{Barzinji:2018xvu}.
The SMEFT extends the SM with higher dimensional operators $\mathcal{Q}_i^{(d)}$
of mass dimension $d$,
\begin{align}
	\Lagr_{\textrm{SMEFT}} &= \Lagr_{\textrm{SM}} + \Lagr^{(5)}+\Lagr^{(6)} +
	\Lagr^{(7)} + \dots  \\ \nonumber \Lagr^{(d)} &= \sum_i \frac{\mathcal{C}_i}{\Lambda^{d-4}}\mathcal{Q}_i^{(d)}
	\textrm{ for } d>4.
\end{align}
The operators are suppressed by $d-4$ powers of the cut-off scale $\Lambda$ and
the $\mathcal{C}_i$ are the Wilson coefficients.
The  $\mathcal{Q}_i^{(d)}$ are constructed out of all of the SM fields and the mass dimension label on the operators is suppressed.
We use the non-redundant Warsaw basis \cite{Grzadkowski:2010es} for $\mathcal{L}^{(6)}$,
which removed some redundancies in the result reported in Ref.~\cite{Buchmuller:1985jz}. (See also Refs.~\cite{AguilarSaavedra:2010zi,Alonso:2014zka}.)

The LEFT is given by
\begin{align}\label{LEFT}
	L_{\textrm{LEFT}} &=L^{\textrm{SM}}_{\textrm{LEFT}} + L^{(5)} + L^{(6)} +
	L^{(7)} + \dots  \\ \nonumber 
	L^{(d)} &= \sum_i \frac{C_i}{\bar{v}_T^{d-4}}\mathcal{P}_i^{(d)} \textrm{ for } d>4,
\end{align}
where
\begin{align}
	L^{\textrm{SM}}_{\textrm{LEFT}} =& -\frac{1}{4}\left[F_{\mu\nu}F^{\mu\nu} + G_{\mu\nu}^AG^{A\mu\nu}\right]
	+ \frac{\theta_{\textrm{QCD}}}{32\pi^2}G_{\mu\nu}^A\tilde{G}^{A\mu\nu}\nonumber \\
	& \hspace{-1.4cm}+ \frac{\theta_{\textrm{QED}}}{32\pi^2}F_{\mu\nu}^A\tilde{F}^{A\mu\nu}
	+ \sum_{\psi} \overline{\psi} i \slashed D \psi  + \overline{\nu}_L i \slashed D \nu_L  + L_{\textrm{LEFT}}^{(3)}.
\end{align}
The dual fields are defined with the convention $\widetilde F_{\mu \nu} =(1/2) \epsilon_{\mu \nu \alpha \beta} F^{\alpha \beta}$
with $\epsilon_{0123}=+1$. The dimension four mass terms are
\bea
- L_{\textrm{LEFT}}^{(3)} =  \sum_{\psi} \overline{\psi}_{\substack{R \\ r}} \! \left[ M_\psi \right]_{rs} \! \! \psi_{\substack{L \\ s}} + \bar{v}_T \, C_{\substack{\nu\\ rs}} \bar{\nu}^c_{\substack{L \\ r}}  \nu_{\substack{L \\ s}} +
		\textrm{h.c.} 
\eea
$F^{\mu \nu} = \partial^\mu A^\nu - \partial^\nu A^\mu$ is the field strength of $\rm U(1)_{em}$.
Here $\psi = \{e,u,d\}$ labels the fermion fields. In the chiral basis for the $\gamma_i$ we use, charge conjugation is given by
$C= - i \gamma_2 \, \gamma_0$. This $C$ is not to be confused with a Wilson coefficient $C_i$.
As chiral projection and charge conjugation do not commute, we fix notation $\psi_L^c = C \, \bar{\psi}_L^T$.
$C_{\substack{\nu}}$ has been rescaled by $\bar{v}_T$ and has mass dimension zero.

The  $\mathcal{P}_i^{(\rm{d})}$ are constructed out of the SM fields except the Higgs, $W^\pm$, $Z$ and the chiral top fields $t_{L,R}$.
The dimensionfull cut off scale of the operators has been chosen to be $\bar{v}_T$ in the LEFT.
The relative couplings required to transform this scale into the mass of a particle integrated out (or a numerical factor in the case of $\Lambda$)
are absorbed by the Wilson coefficients.

{\bf Equations of Motion.} The SM, the SMEFT and the LEFT are all consistent field theories defined by actions
\bea
S = \int \mathcal{L}(\chi, \partial \chi) d^{4-2 \epsilon} x.
\eea
Each theory contains field variables, here generically denoted $\chi$. The meaning of the field variables, even those with the same notational label,
differs in these theories. A field is redefined order by order in an EFT power counting expansion to remove redundancies of description out of the Lagrangian.
As a result, the extremum of the action under variations of field configurations,
\bea\label{eulerlagrange}
0 = \delta S = \int d^{4-2 \epsilon} x \left[\frac{\partial \mathcal{L}}{\partial \chi} \delta \chi - \partial_\mu \left(\frac{\partial \mathcal{L}}{\partial (\partial_\mu \chi)}\right) \,  \delta \chi\right],
\eea
is also redefined order by order. The descendent EOM for $\chi$ then
depend on the local contact operators that are present in the EFT expansion. Asymptotic states  can be considered
to be free field solutions to the modified EOM. The $\Delta$ corrections to the EOM  modify matching to sub-leading order onto an EFT \cite{Elgaard-Clausen:2017xkq,Jenkins:2017dyc,Barzinji:2018xvu}, 
and modify the sources of gauge fields. 
Obviously, one must be careful to include all effects when dealing with higher orders in the 
power counting expansion.

For the LEFT  the gauge fields have the expanded EOM
\begin{align}
	D_{\nu}F^{\nu\mu} =& e\sum_\psi \overline{\psi}Q\gamma^{\mu}\psi
	+ 4 \frac{\theta_{\textrm{QED}}}{32\pi^2}\partial_{\nu}\tilde{F}^{\nu\mu}
	+ \sum_{d} \frac{\Delta^{\mu,(d)}_{F}}{\bar{v}_T^{d-4}}, \\
	\left[D_{\nu},G^{\nu\mu}\right]^A =& g_3\sum \overline{\psi}\gamma^{\mu}T^A\psi
	+ 4 \frac{\theta_{\textrm{QCD}}}{32\pi^2}\left[D_{\nu},\tilde{G}^{\nu\mu}\right]^A
	\nonumber \\
	&+ \sum_{d} \frac{\Delta^{A\mu,(d)}_{G}}{\bar{v}_T^{d-4}}.
\end{align}
Here we have used the adjoint derivative with definition
\bea
\left[D^\alpha , \mathcal{Q} \right]^A = \partial^\alpha \, \mathcal{Q}^A - g_3 \, f^{BCA} \, G^{\alpha}_B \mathcal{Q}_C.
\eea
For the fermions, the EOM take the form
\begin{align}
	i \slashed D \psi_{\substack{R \\ p}} &= \left[M_{\psi}\right]_{pr}\psi_{\substack{L \\ r}} - \sum_{d=5}^\infty \frac{\Delta^{(d)}_{\psi_R,p}}{\bar{v}_T^{d-4}}, \\
	i \slashed D \nu_{\substack{L \\ p}} &= - \sum_{d=3}^\infty \frac{\Delta^{(d)}_{\nu_L,p}}{\bar{v}_T^{d-4}}, \\
	i \slashed D \psi_{\substack{L \\ p}} &= \left[M_{\psi}^\dagger\right]_{pr} \psi_{\substack{R \\ r}} - \sum_{d=5}^\infty \frac{\Delta^{(d)}_{\psi_L,p}}{\bar{v}_T^{d-4}}.
\end{align}
Each $\Delta^{(d)}$ up to $L_{\rm LEFT}^{(6)}$ is given in the Appendix.

{\bf Symmetry Currents.} A continuous transformation of a field,
\begin{align}
	 \chi(x)  \rightarrow  \chi'(x)  = \chi(x)  + \alpha \nabla  \chi(x),
\end{align}
under a deformation $ \nabla  \chi(x)$, with an associated infinitesimal parameter $\alpha$,
is a symmetry of $S$ if $S \rightarrow S'$ is invariant under this transformation, up to the possible generation of a surface term.
The EOM defined by the variations of field configurations in the action --$\delta S$-- is unchanged by this transformation.
The EOM are defined with surface terms neglected, and
the surface terms themselves are defined to be those of the form
\begin{align}
\partial_\mu \left(\frac{\partial \mathcal{L}}{\partial (\partial_\mu \chi)} \, \nabla \chi \right),
\end{align}
generated by $\delta S$.
The Lagrangian is then invariant under $S \rightarrow S'$,  up to a possible total derivative
\begin{align}
	\label{eq:lagrtransform}
	\Lagr \rightarrow \Lagr + \alpha \partial_\mu \mathcal{K}^\mu,
\end{align}
for some $\mathcal{K}^\mu$. Associated with each symmetry defined in this manner is a conserved current  \cite{Noether1918}.
The definition of the current is
\begin{align}
	J^\mu = \frac{\partial \Lagr}{\partial \left(\partial_\mu \chi \right)}\nabla \chi- \mathcal{K}^\mu.
\end{align}
The conservation of the current corresponds to
\begin{align}
\partial_\mu J^\mu = 0.
\end{align}
Due to the presence of an EFT power counting expansion, it is interesting to examine
how symmetry currents are defined when non-renormalizable operators are included, and how these currents encode symmetry constraints.

{\bf Basis dependence.}
The symmetry currents are basis dependent in an EFT, but still meaningful.
They receive corrections due to the local contact operators in a particular basis through the modification of the EOM.
The basis dependence of the symmetry currents can be made clear by considering a space-time symmetry. For an infinitesimal translation of this form
\begin{align}
	x^\mu &\rightarrow x^\mu - a^\mu, \nonumber \\
	\chi(x) &\rightarrow \chi(x+a) = \chi(x) + a^\mu\partial_\mu\chi(x), \nonumber \\
	\Lagr &\rightarrow \Lagr + a^\mu\partial_\mu \Lagr = \Lagr + a^\nu\partial_\mu\left(\delta^\mu_\nu
	\Lagr\right),
\end{align}
up to $\mathcal{O}(a^2)$.
Comparing to Eqn.~\eqref{eq:lagrtransform} identifies $\mathcal{K}$.
Four separately conserved currents result, identified as 
the stress-energy tensor, given by
\begin{align}
	T^\mu_\nu = \frac{\partial \Lagr}{\partial \left(\partial_\mu\chi\right)}\partial_\nu\chi
	- \Lagr\delta^\mu_\nu.
\end{align}
The $\chi$ become basis dependent
when redundant operators are removed from the EFT, leading to the chosen basis of operators for $\Lagr$. The $T^\mu_\nu$ constructed from $\{\chi, \Lagr\}$ is also
basis dependent as a result at the same order in the power counting. This should be unsurprising,
as the currents are auxiliary operators, and sources and the related Green's functions are not invariant
under field redefinitions.
For more detailed discussion on this point, see Refs.~\cite{Politzer:1980me,Brivio:2017vri}.
This basis dependence is similar to scheme dependence. It vanishes in relationships between
a set of physical measured quantities (i.e. $S$-matrix elements constructed with an LSZ procedure) defined via the same stress-energy tensor.
Symmetry constraints between $S$-matrix elements are basis independent, even though the symmetry current itself carries basis dependence.

{\bf Non-linear global symmetries.}
The effect of non-linear representations of the symmetries of the LEFT is straightforward in some cases.
As a simple example, consider transforming the charged lepton fields as
\begin{align}\label{chargedlepton}
	e_{\substack{L \\p}} &\rightarrow e^{i\alpha}e_{\substack{L \\p}}, & \quad
	e_{\substack{R \\p}} &\rightarrow e^{i\alpha}e_{\substack{R \\p}},
\end{align}
by some global phase $\alpha$. By inspection of the LEFT operator basis, the $\Delta L = 0$ operators all respect this transformation, except $\mathcal{O}_{\nu edu}$.
The charged lepton current is
\begin{align}
	J_{\substack{e \\rr}}^\mu \equiv J_{\substack{e,L \\rr}}^\mu + J_{\substack{e,R \\rr}}^\mu 
\equiv\overline e_{\substack{L \\r}} \gamma^\mu e_{\substack{L \\r}} + \overline e_{\substack{R \\r}} \gamma^\mu e_{\substack{R \\r}} + \dots
\end{align}
The kinetic terms are taken to a flavour diagonal form
\begin{align}
	\psi_{\substack{L/R \\ r}} \rightarrow U(\psi, L/R)_{rs} \psi'_{\substack{L/R \\ s}},
\end{align}
using the flavour space rotation matrix $U$. In the remainder of the paper, the prime superscript is suppressed.
$J_{e}$ descends from the kinetic terms and is also flavour diagonal after these rotations. $J_{e}$ can receive contributions from higher
dimensional operators in a basis, as indicated by the ellipsis in the above expression.
The LEFT basis of Refs.~\cite{Jenkins:2017jig,Jenkins:2017dyc} removes derivative operators systematically so there are no
contributions of this form due to the $L_{\rm LEFT}^{(6)}$ defined in these works. The divergence of the current including the EOM corrections $\Delta^{(6)}$ is
\begin{align}
	\label{eq:partialje}
	i \partial_\mu J_{\substack{e,L\\ rr}}^\mu =& i\left(\partial_\mu \overline e_{\substack{L \\r}} \right) \gamma^\mu e_{\substack{L \\r}}
	+ i\overline e_{\substack{L \\p}}\gamma^\mu\left(\partial_\mu e_{\substack{L \\p}}\right)  \\
	=& \left(-\overline e_{\substack{R \\p}} M_{\substack{e \\pr}} + \Delta^{(6)}_{\overline e_{\substack{L \\r}}}\right)e_{\substack{L \\r}}
	+ \overline e_{\substack{L \\p}}\left( M_{\substack{e \\rp}}e_{\substack{R \\r}} - \Delta^{(6)}_{e_{\substack{L \\p}}}\right),  \nonumber 
\end{align}
and similarly for $i \partial_\mu J_{\substack{e,R}}^\mu$.
The mass terms are invariant under Eqn.~(\ref{chargedlepton}) and cancel when the expressions are summed. 
We split the EOM correction and $J$ into lepton number conserving and violating parts,
$\Delta^{(6)} = \Delta^{(6,L)}+\Delta^{(6,\slashed L)}$ and $J^\mu = J^{(L)\mu} + J^{(\slashed L)\mu}$.
First, consider the lepton number conserving part of Eqn.~\eqref{eq:partialje}. A significant degree of cancellation occurs in the resulting expression. The
only Wilson coefficient remaining corresponds to $\mathcal{P}_{\nu edu}$, an operator which is not individually invariant under the charged lepton field transformation. The explicit expression is
\begin{widetext}
\bea
	\Delta^{(6,L)}_{\overline e_{\substack{L \\r}}} e_{\substack{L \\r}} - \overline e_{\substack{L \\p}}\Delta^{(6,L)}_{e_{\substack{L \\p}}}
	&+& \Delta^{(6,L)}_{\overline e_{\substack{R \\r}}} e_{\substack{R \\r}} - \overline e_{\substack{R \\p}}\Delta^{(6,L)}_{e_{\substack{R \\p}}}
	=\left(C^{V,LL}_{\substack{\nu edu \\ prst}}J_{\substack{\nu e,L \\ pr}}^{\mu}J_{\substack{du,L \\ st}}^{\nu}
	- C^{V,LL*}_{\substack{\nu edu \\ rpts}}J_{\substack{e\nu,L \\ pr}}^{\mu}J_{\substack{ud,L \\ st}}^{\nu}\right)\eta_{\mu\nu}
	\nonumber \\
	&+&\left( C^{V,LR}_{\substack{\nu edu \\ prst}}J_{\substack{\nu e,L \\ pr}}^{\mu}J_{\substack{du,R \\ st}}^{\nu}
	- C^{V,LR*}_{\substack{\nu edu \\ rpts}} J_{\substack{e\nu,L \\ pr}}^{\mu}J_{\substack{ud,R \\ st}}^{\nu}\right)\eta_{\mu\nu}
	+ C^{S,RR}_{\substack{\nu edu \\ prst}}S_{\substack{\nu e,L \\ pr}}S_{\substack{du,L \\ st}}
	- C^{S,RR*}_{\substack{\nu edu \\ rpts}}S_{\substack{e\nu,R \\ pr}}S_{\substack{ud,R \\ st}}
	 \\
	&+&\left( C^{T,RR}_{\substack{\nu edu \\ prst}}\cT_{\substack{\nu e,L \\ pr}}^{\mu\nu}\cT_{\substack{du,L \\ st}}^{\alpha\beta}
	- C^{T,RR*}_{\substack{\nu edu\\rpts}}\cT_{\substack{e\nu , R \\ pr}}^{\mu\nu}\cT_{\substack{ud,R \\ st}}^{\alpha\beta}\right)\eta_{\alpha\mu}\eta_{\beta\nu}
	+ C^{S,RL}_{\substack{\nu edu\\ prst}}S_{\substack{\nu e,L \\ pr}}S_{\substack{du,R \\ st}}
	- C^{S,RL*}_{\substack{\nu edu\\ rpts}}S_{\substack{e\nu,R \\ pr}}S_{\substack{ud,L \\ st}}. \nonumber
\eea
\end{widetext}
Similarly, we can define a neutrino current
\begin{align}
	J_{\substack{\nu \\ rr}}^\mu \equiv \overline \nu_{\substack{L\\r}}\gamma^\mu\nu_{\substack{L\\r}} + \dots
\end{align}
The lepton number conserving contributions to the divergence of the neutrino current are such that
\begin{align}
	\Delta^{(6,L)}_{\overline e_{\substack{L \\r}}} e_{\substack{L \\r}} - \overline e_{\substack{L \\p}}\Delta^{(6,L)}_{e_{\substack{L \\p}}}
		&+\Delta^{(6,L)}_{\overline e_{\substack{R \\r}}} e_{\substack{R \\r}} - \overline e_{\substack{R \\p}}\Delta^{(6,L)}_{e_{\substack{R \\p}}} \nonumber \\
		&+\Delta^{(6,L)}_{\overline \nu_{\substack{L \\r}}} \nu_{\substack{L \\r}} - \overline \nu_{\substack{L \\p}}\Delta^{(6,L)}_{\nu_{\substack{L \\p}}}
	=0.
\end{align}
This is as expected, and provides a cross check of the EOM corrections in the Appendix. The total lepton field current is conserved by the subset of $\Delta L = 0$ operators leading to
\begin{align}
	\partial_\mu J_\ell^{(L)\mu} = 0,
\end{align}
where $\ell$ is the $\rm SU(2)_L$ doublet field. Considering the transformation of only part of the lepton multiplet under a phase change also illustrates
how a symmetry can be present in a Lagrangian, but non-linearly realized. The symmetry constraint is only made manifest when all terms corresponding to the linear symmetry multiplet are simultaneously
included in the constructed symmetry current.
This re-emphasizes the requirement to use a consistent LEFT with all operators retained when studying the data. Doing so ensures that the LEFT represents a consistent IR limit.
Conversely, dropping operators can forbid non-linear realizations in the LEFT of UV symmetries, which can block a consistent IR limit of some UV completions being defined.
For this reason (see also Ref.~\cite{Jiang:2016czg}), experimental studies of constraints on higher dimensional operators done ``one at a time" can result in misleading conclusions.

{\bf Linear representations of global symmetries.}
Operator dimension in the SMEFT is even (odd) if
$(\Delta B - \Delta L)/2$ is even (odd)~\cite{deGouvea:2014lva,Kobach:2016ami}.
Here $\Delta B$ and $\Delta L$ are respectively the baryon and lepton number violation of the operator
considered. In $\mathcal{L}_{\rm SM} + \mathcal{L}^{(6)}$, $B-L$ is an accidentally conserved quantity consistent with this constraint.

In the LEFT, incomplete $\rm SU(2)_L$ SM multiplets are used to construct operators, and operators are not constructed to respect
hypercharge. The relationship between operator dimension and global lepton and baryon number in the LEFT is different
than in the SMEFT as a result.
When considering arbitrary Wilson coefficients in the LEFT, the classes of $\Delta L=2$, $\Delta B= - \Delta L= 1$, and $\Delta L = 4$
 defined in Refs.~\cite{Jenkins:2017jig,Jenkins:2017dyc} are present.
These $\psi^4$ operators are not present in $ \mathcal{L}^{(6)}$ in the SMEFT, and
these operators violate $B-L$.

The SMEFT relationship between operator dimension and these global symmetries is projected onto the LEFT operator basis
when the matching result of Ref.~\cite{Jenkins:2017jig} is imposed. The corresponding $\mathcal{L}_{\rm{SMEFT}}$ - $L_{\rm{LEFT}}$
matchings that violate $B-L$ vanish exactly.

{\bf Hypercharge.}
The fermion hypercharge current of the SM is
 \begin{align}
	J_{ \Psi \hyp,\rm SM}^\mu = \hspace{-0.4cm} \sum_{\substack{\Psi = e_R,u_R,d_R,\\ \ell_L,q_L}}  \hspace{-0.4cm}\hyp_\Psi \overline\Psi  \gamma^\mu \Psi,
\end{align}
where $\hyp_\Psi = \{-1,2/3,-1/3,-1/2,1/6\}$.
This current is manifestly not conserved in the LEFT
\begin{align}
	\partial_\mu J_{\Psi \hyp,\rm SM}^\mu \neq 0.
\end{align}
In the LEFT, a hypercharge current can be defined as
 \begin{align}
	J_{ \Upsilon \hyp}^\mu =\sum_{\substack{\Upsilon}} \hyp_\Upsilon \overline \Upsilon  \gamma^\mu \Upsilon.
\end{align}
Here $\Upsilon = \{\psi_R, \psi_L, \nu_L \}$  and the hypercharges are assigned as in the SM.
Part of the non-conservation of the current stems from the fermion mass terms.  In addition, the $\Delta$ corrections also lead to the current not being conserved when the Wilson
coefficients in the LEFT take arbitrary values.
When the matching conditions on the Wilson coefficients to the SMEFT are imposed \cite{Jenkins:2017jig}, many of the EOM corrections generating a non-vanishing $\partial_\mu J_{\Upsilon \hyp}^\mu$ are removed. 
The terms that remain are
\begin{widetext}
\begin{align}\label{hypercharge}
	i\partial_\mu J_{\Upsilon \hyp}^\mu \Bigr\rvert_\textrm{match} &=
	\frac{( \hyp_{u_R} - \hyp_{d_R})}{\bar{v}_T^2}\left(\underset{prst}{C^{V,LR}_{\nu edu}} J_{\substack{\nu e,L \\ pr}}^{\mu}J_{\substack{du,R \\ st}}^{\nu}
	-\underset{rpts}{C^{V,LR*}_{\nu edu}}J_{\substack{e\nu,L \\ pr}}^{\mu}J_{\substack{ud,R \\ st}}^{\nu}
+\underset{prst}{C^{V1,LR}_{uddu}}J_{\substack{ud,L \\ pr}}^{\mu}J_{\substack{du,R \\ st}}^{\nu}
-\underset{rpts}{C^{V1,LR*}_{uddu}}J_{\substack{du,L \\ pr}}^{\mu}J_{\substack{du,R \\ st}}^{\nu}\right)\eta_{\mu\nu} \nonumber  \\
&+ ( \hyp_{\psi_R} - \hyp_{\psi_L})\left( \overline{\psi}_{\substack{R \\ p}} \left[M_{\psi}\right]_{pr} \! \!  \psi_{\substack{L \\ r}}-
 \overline{\psi}_{\substack{L \\ p}} \left[M_{\psi}^\dagger \right]_{pr} \! \!  \psi_{\substack{R \\ r}}\right)
+ 2 \, \bar{v}_T \hyp_{\nu_L} \left[\overline{\nu}_{\substack{L\\p}} C_{\substack{\nu \\ pr}}^\star \nu_{\substack{L\\r}}^c
- \overline{\nu}^c_{\substack{L\\p}} C_{\substack{\nu \\ pr}}^T \nu_{\substack{L\\r}}\right]\\
&+ \frac{(\hyp_{\psi_L} - \hyp_{\psi_R})}{\bar{v}_T}  \sum_{\psi \neq e} \left[\overline{\psi}_{\substack{R\\p}} \sigma^{\alpha \beta} T^A \psi_{\substack{L\\r}}  \, C_{\substack{\psi G \\ rp}}^\star
 - \overline{\psi}_{\substack{L\\p}} \sigma^{\alpha \beta} T^A \psi_{\substack{R\\r}} \, C_{\substack{\psi G\\ rp}}^T \right] G_A^{\alpha \beta} \nonumber \\
 & + \frac{(\hyp_{\psi_L} - \hyp_{\psi_R})}{\bar{v}_T} \left[\overline{\psi}_{\substack{R\\p}} \sigma^{\alpha \beta} \psi_{\substack{L\\r}} \, C_{\substack{\psi \gamma \\ rp}}^\star
- \overline{\psi}_{\substack{L\\p}} \sigma^{\alpha \beta} \psi_{\substack{R\\r}}  \, C_{\substack{\psi \gamma \\ rp}}^T \right] F_{\alpha \beta} + \dots \nonumber
\end{align}
\end{widetext}
Here we have used the fact that in whole or in part, composite operators forms with $\sum_\Psi  \hyp_\Psi = 0$ have a corresponding vanishing contribution
to the current. This condition being fulfilled also provides a cross check of the $\Delta^{(3 -6)}$ EOM corrections in the Appendix.

Enforcing matching constraints to the SMEFT is insufficient to make the hypercharge current manifest.
The reason is that SM states are integrated out in constructing the LEFT, that carry this quantum number.
Consider the definition of the full hypercharge current
\begin{align}
	J_{{\hyp},\textrm{full}}^{\mu} = J_{ \Psi \hyp}^\mu
	+  \hyp_H H^\dagger \, i\overleftrightarrow D^\mu H,
\end{align}
where $\hyp_H=1/2$ for the Higgs field. Here, and later, we are using the Hermitian derivative defined by
\bea\label{hermitian}
O^\dagger \, i\overleftrightarrow D_\mu O &=& i O^\dagger (D_\mu O) - i (D_\mu O)^\dagger O,
\\
	O^\dagger i \overleftrightarrow{D}^I_\mu O &=& i O^\dagger \tau^I \left( D_\mu O\right)
	- i \left( D_\mu O\right)^\dagger \tau^I O,
\eea
for a field $O$. To make hypercharge conservation manifest, we include the transformation properties of the masses associated with states integrated out that depended on $\langle H^\dagger H \rangle$.
This can be done in a spurion analysis. Rescaled Wilson coefficients and mass terms are promoted to spurion fields  with tilde superscripts
\begin{align*}
\tilde C^{V,LR}_{\substack{\nu edu \\ prst}}&=\bar{v}_T C^{V,LR}_{\substack{\nu edu\\ prst}}, & \quad
\tilde C^{V1,LR}_{\substack{uddu\\ prst}}&=\bar{v}_T C^{V1,LR}_{\substack{uddu\\prst}}, \\
\tilde C_{\substack{\psi \\ pr}} &= M_{\substack{\psi\\ pr}}, & \quad \quad \tilde C_{\substack{\nu \\ pr}} &=  2 \, \bar{v}_T \, C^\star_{\substack{\nu \\ pr}}, \\
\tilde C_{\substack{\psi \gamma \\ pr}} &= \bar{v}_T \, C_{\substack{\psi \gamma \\ rp}}^\star,
& \quad \quad \tilde C_{\substack{\psi G \\ pr}} &= \bar{v}_T \, C_{\substack{\psi G \\ rp}}^\star.
\end{align*}
These spurion fields have the hypercharge assignments
\begin{align*}
\hyp_{\tilde C} &=\hyp_{d_R} -\hyp_{u_R} & \, \,  &\text{for} \, \,  \, \tilde C^{V,LR}_{\substack{\nu edu}}, \, \, \, \tilde C^{V1,LR}_{\substack{uddu}}, \\
\hyp_{\tilde C} &= - \hyp_\nu & \, \,  &\text{for} \, \,  \, \tilde C_\nu, \\
\hyp_{\tilde C} &= \hyp_{\psi_R} - \hyp_{\psi_L} & \, \, &\text{for} \, \,  \, \tilde C_{\substack{\psi \gamma}}, \tilde C_{\substack{\psi G }}, \\
\hyp_{\tilde C} &= \hyp_{\psi_L} - \hyp_{\psi_R} & \, \, &\text{for} \, \,  \, \tilde C_{\substack{\psi}}.
\end{align*}
As the spurions are charged under hypercharge, we need to include them in the
current in the LEFT
\begin{align}
	J_{\hyp,\textrm{LEFT}}^\mu =J_{ \Upsilon \hyp}^\mu + J_{\hyp,\textrm{S}}^\mu,
\end{align}
where
\begin{align}
	J_{\hyp,\textrm{S}}^\mu =&  \sum_{\tilde{C}} \hyp_{\tilde C} \, \tilde C^\dagger \, i \overleftrightarrow D^\mu \tilde C.
\end{align}
Here the flavour indices are suppressed.
When promoting the Wilson coefficients to fields, we need to include kinetic terms,

\begin{align}
	\Lagr_{\textrm{S}}^{kin} =&\sum_{\tilde{C}} \left(D^\mu \tilde C\right)^\dagger \left(D_\mu \tilde C\right).
\end{align}
The EOM for the spurion fields are $ D^2 \tilde C = \delta L_{\rm LEFT}/\delta \tilde C^\star$.
Including these contributions, the hypercharge current is conserved: $ i\partial_\mu J_{\hyp,\textrm{LEFT}}^\mu = 0$.

This provides a cross check of the EOM corrections in the Appendix and the results in Ref.~\cite{Barzinji:2018xvu,Jenkins:2017jig}.

{\bf $\rm SU(2)_L$ current.}
The $\rm SU(2)_L$ current in the SMEFT is defined as
\begin{align}
	\label{eq:su2current}
	J_\mu^I = \frac{1}{2} \overline{q}\tau^I\gamma_\mu q
	+ \frac{1}{2}\overline{l}\tau^I \gamma_\mu l
	+\frac{1}{2}H^\dagger i \overleftrightarrow{D}^I_\mu H.
\end{align}
This definition of the current fixes the embedding of the LEFT states into $\rm SU(2)_L$ doublets.
Here $\tau^I$ are the $\rm SU(2)_L$ generators (Pauli matrices) with normalization $[\tau^I,\tau^J] = 2 \, i \, \epsilon_{IJK} \tau^K$ for $I = \{1,2,3\}$.
The fields $q$ and $l$ are left-handed quark and lepton $\rm SU(2)_L$ doublets, which are absent in the LEFT as linear multiplets.
To examine the  $\rm SU(2)_L$ current we need to combine terms in the LEFT into reconstructed
 $\rm SU(2)_L$  multiplets and also introduce spurions to account for the transformation properties of $\bar{v}_T$.
We illustrate the constraints of the $\rm SU(2)_L$ current with an operator
from the class $(\overline L R)X + \textrm{h.c.}$ as an example,
\begin{align}
	C_{\substack{e\gamma\\ pr}} \overline e_{\substack{L \\ p}} \sigma^{\mu\nu}
	e_{\substack{R \\ r}} F_{\mu\nu} + \textrm{h.c.}
	\rightarrow
	\overline{l}^i_{\substack{L \\ p}} \sigma^{\mu\nu} e_{\substack{R \\ r}}
	F_{\mu\nu} C^i_{\substack{e\gamma\\ pr}} + \textrm{h.c.}
\end{align}
where
\begin{align}
	\underset{pr}{C^i_{e\gamma}} = \begin{pmatrix}
		0 \\
		C_{e\gamma}
	\end{pmatrix}_{pr}
	\qquad \textrm{and}
	\qquad
	l^i_{\substack{L \\ p}} = \begin{pmatrix}
		\nu_{\substack{L \\ p}} \\
		e_{\substack{L \\ p}}
	\end{pmatrix}.
\end{align}
We have promoted the Wilson coefficient to a $\rm SU(2)_L$ doublet field, and
collected the left-handed leptons into a doublet.
Analogous promotions can be made for all the operators in
this class. The
relevant terms in the equations of motion are
\begin{align}
	\bar{v}_T \, i\slashed D l^i_{\substack{L \\ p}} &= - \sigma^{\mu\nu}e_{\substack{R \\ r}} F_{\mu\nu}
	C^i_{\substack{e\gamma\\ pr}} +\dots \\
	\bar{v}_T \, i\slashed D \overline l^i_{\substack{L \\ p}} &= + C^{i*}_{\substack{e\gamma\\ pr}}F_{\mu\nu}\overline e_{\substack{R \\ r}}\sigma^{\mu\nu} + \dots \\
	\label{eq:su2spurion1}
	\bar{v}_T^2 \, D^2 \tilde C^i_{\substack{e\gamma\\ pr}} &= F_{\mu\nu}\overline e_{\substack{R \\ r}} \sigma^{\mu\nu} l^i_{\substack{L \\ p}}, \\
	\label{eq:su2spurion2}
	\bar{v}_T^2 \, D^2 \tilde C^{i*}_{\substack{e\gamma\\ pr}} &= \overline{l}^i_{\substack{L \\ p}}\sigma^{\mu\nu} e_{\substack{R \\ r}} F_{\mu\nu}.
\end{align}
The covariant derivative of $J_{l}^{\mu}$ gives
\bea
\label{eq:su2currentdiv}
i \left[D_\mu, J_{l}^{\mu}\right]^I &\equiv& i\partial_{\mu}\left( \frac{1}{2} \overline l_{\substack{L\\p}}
\tau^I\gamma^\mu l_{\substack{L \\ p}}\right)  - g_2 \epsilon^{JKI}W_{\mu,J} J_{l,K}^\mu\\
&=&\frac{1}{2} \left( iD_\mu \overline l_{\substack{L \\ p}}\right) \tau^I \gamma^\mu l_{\substack{L \\ p}}
+\frac{1}{2} \overline l_{\substack{L \\ p}} \tau^I \gamma^\mu \left(i D_\mu l_{\substack{L \\ p}}  \right) \nonumber \\
&=& \frac{ C_{\substack{e\gamma\\ pr}}^{*}}{2 \bar{v}_T} F_{\mu\nu} \overline e_{\substack{R \\ r}} \sigma^{\mu\nu} \tau^I l_{\substack{L \\ p}}
- \frac{C_{\substack{e\gamma\\ pr}}}{2 \bar{v}_T} \overline l_{\substack{L \\ p}} \tau^I \sigma^{\mu\nu} e_{\substack{R \\ r}} F_{\mu\nu}  + \dots  \nonumber
\eea
To recover a conserved current, we perform a spurion analysis, similar to the
one for hypercharge. We have the EOM for the spurion $C_{e\gamma}$,
in Eqns.~\eqref{eq:su2spurion1} and~\eqref{eq:su2spurion2}.
The spurion current is
\bea
J_{S}^{\mu,I} =
\frac{1}{2} \tilde C_{\substack{e\gamma}}^\dagger i \overleftrightarrow{D}^{\mu,I} \tilde C_{\substack{e\gamma}},
\eea
with flavour indices suppressed. The covariant divergence of the spurion current is
\bea
\label{eq:su2spuriondiv}
i \left[D_\mu, J_S^{\mu}\right]^I &=& -\frac{1}{2}\left[\underset{pr}{ \tilde C_{e\gamma}}^*\tau^I D^2\underset{pr}{ \tilde C_{e\gamma}}
- D^2\underset{pr}{ \tilde C_{e\gamma}}^* \tau^I \underset{pr}{ \tilde C_{e\gamma}}\right] \\
&\,& \hspace{-2cm} =-\frac{1}{2 \bar{v}_T}\left[\underset{pr}{C_{e\gamma}}^* \tau^I F_{\mu\nu}\overline e_{\substack{R \\ r}} \sigma^{\mu\nu} l_{\substack{L \\ p}}
- \overline l_{\substack{L \\ p}} \sigma^{\mu\nu} e_{\substack{R \\ r}} F_{\mu\nu} \tau^I \underset{pr}{C_{e\gamma}}\right].  \nonumber
\eea
Combining Eqns.~\eqref{eq:su2currentdiv} and~\eqref{eq:su2spuriondiv}, the new current is
covariantly conserved for the chosen operator from the class $(\overline L R) X$,
\bea
i\left[D_{\mu}, J^{\mu}\right]^I \equiv i\left[D_{\mu}, \left( J^{\mu}_l+ J^{\mu}_S\right)\right]^I = 0.
\eea
The generalization to include quarks is straightforward.

For $\psi^4$ operators a similar spurion analysis that also includes the
promotion of all of the fermion fields into the corresponding $\rm SU(2)_L$ fermion multiplet of the
SM is done. The procedure is straightforward. When imposing the $\mathcal{L}_{\rm{SMEFT}}$ - $L_{\rm{LEFT}}$ matching
and performing this spurion analysis, the $\rm SU(2)_L$ current is conserved.

{\bf Contraints due to non-manifest currents.}
The $\rm SU(2)_L$ and $\rm U(1)_Y$ currents are not conserved in the LEFT when the Wilson coefficients of this theory are treated
as free parameters. Furthermore, the implication of these currents in the LEFT is distinct than in the SM or the SMEFT,
as there is no manifest field corresponding to these currents when they are conserved. There is no direct construction of
a Ward identity using a propagating gauge field as a result.

The conserved currents do constrain the LEFT by fixing relationships between otherwise free parameters of the theory.
Matrix elements of the currents can be directly constructed, as they are composed
of the fields of the LEFT. Constructing such a matrix element from the generalized Heisenberg current field,
with a set of initial and final states denoted $\Psi_{i,f}$, and taking a total derivative gives
\bea
\partial^{\mu} \int d^{4} x e^{i p \cdot x}  \langle \Psi_f | J_\mu(x) |\Psi_i \rangle = 0.
\eea
A series of relationships between the Wilson coefficients then follows
\bea
\sum_n \partial_\mu \langle \Psi_f | \mathcal{P}_n | \Psi_i \rangle^{\mu}(p) C_n = 0.
\eea
Formally, the measured $S$-matrix elements must be constructed using an
LSZ reduction formula.
The constraints that follow for the Wilson coefficients are trivially satisfied only if the Wilson coefficients
are already fixed by a UV matching preserving the corresponding symmetry.

{\bf $\rm U(1)_{em}$ and the LEFT multipole expansion.}
The classical limit of $L^{d \leq 4}_{\rm LEFT}$ reproduces the well known physics of Maxwell's equations, and in particular Gauss's law
\cite{lagrange} (see also Ref.~\cite{gauss}).
Gauss's law relates the time component of the  electromagnetic current $J^\mu = \overline \psi_e  \gamma^\mu \psi_e$ to
\bea
J^0 = \frac{\nabla \cdot {\bf E}}{- e_{phys}}.
\eea
Here $e_{phys} = 1.6021766208(98) \times 10^{-19} \, C$, is the electron charge in the usual SI units \cite{RevModPhys.88.035009}.
In the LEFT, the electromagnetic current is also expected to be conserved
\bea
\partial_\mu \, J^\mu =0,
\eea
without any of the subtleties of the previous sections as the $\mathcal{P}_i$ are constructed to manifestly preserve $\rm U(1)_{em}$.

The $\rm U(1)_{em}$ current is subject to its own set of subtleties.
First, the naive understanding that $J^\mu $ being conserved directly leads to its non-renormalization requires some
refinement. This issue was comprehensively addressed
for QED in Ref.~\cite{Collins:2005nj}, neglecting higher dimensional operators and considering a one electron state and
the corresponding electron number current. Here we review the result of 
Refs.~\cite{Collins:2005nj,Manohar:2017eqh} and then directly extend this result into the LEFT.

The definition of the electromagnetic current is affected by
the presence of a surface term $\partial^\nu F_{\nu \mu}$ \cite{lurie,Collins:2005nj} introducing a renormalization of this current.
We define the $L^{d \leq 4}_{\rm LEFT}$ CP conserving QED Lagrangian as
\begin{align*}
\mathcal{L} &= \left(\overline \psi\left[ i \gamma \cdot ( \partial + e q A) - m \right] \psi\right)^{(0)} - \frac{1}{4}(F^{(0)}_{\mu \nu})^2 - \frac{1}{2 \xi}(\partial \cdot A)^2 \\
& \hspace{-0.3cm}= Z_2 \overline \psi \left[ i \gamma \cdot ( \partial + e q \mu^\epsilon A) - m^{(0)} \right] \psi - \frac{Z_3}{4}(F_{\mu \nu})^2 - \frac{1}{2 \xi}(\partial \cdot A)^2,
\end{align*}
where all $()^{(0)}$ superscripted quantities are bare parameters. $\mu$ is introduced so that the renormalized coupling is dimensionless and $q$ is the charge
of $\psi$. We restrict our attention to $\psi = \psi_e$ for simplicity (even in loops) in the discussion below. Renormalized quantities
are introduced above with a suppressed $r$ superscript, $d = 4 - 2 \epsilon$ and we use  $\rm \overline{MS}$ as a subtraction scheme so that
\begin{align*}
A_\mu^{(0)} &= \sqrt{Z_3} A_\mu^{(r)}, \quad
\psi^{(0)} = \sqrt{Z_2} \psi^{(r)}, \\
m_e^{(0)} &= Z_m m_e^{(r)},  \quad
e^{(0)} = Z_e \mu^\epsilon \, e^{(r)}. 
\end{align*}
Here $m_e^2 = [M_e]_{11} [M_e^\dagger]_{11}$.
The renormalization constants in QED are given by
\begin{align*}
Z_3 &= 1 - \frac{e^2 S_\epsilon}{12 \, \pi^2 \, \epsilon}, \quad 
Z_2 = 1 - \frac{e^2 S_\epsilon}{16 \, \pi^2 \, \epsilon}, \\
Z_m &= 1 - \frac{3 e^2 S_\epsilon}{16 \, \pi^2 \, \epsilon},
\end{align*}
and $Z_e = 1/\sqrt{Z_3}$ at one loop. Here $S_\epsilon = (4 \pi e^{-\gamma_E})^\epsilon$, following the notation of Ref.~\cite{Collins:2005nj}.
Hereon we define our subtractions in $\rm \overline{MS}$ and suppress the corresponding constant terms, setting $S_\epsilon = 1$.

\begin{figure}[!ht]
\includegraphics[width=0.2\textwidth]{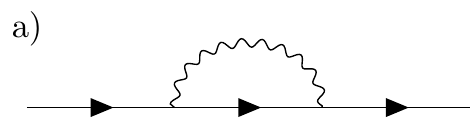}
\includegraphics[width=0.2\textwidth]{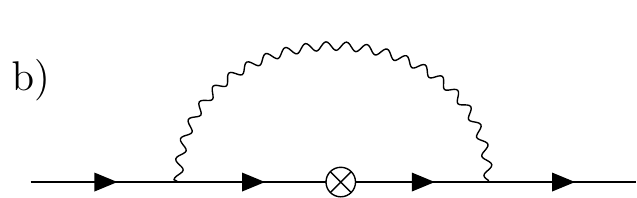}
\linebreak \\
\includegraphics[width=0.15\textwidth]{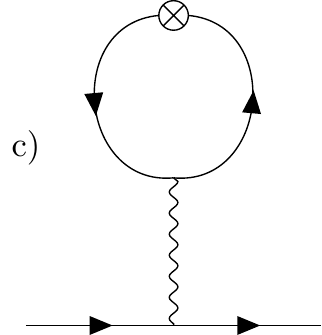}
\includegraphics[width=0.15\textwidth]{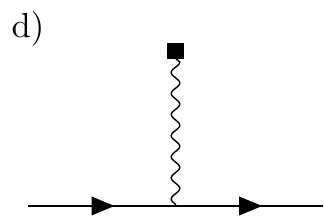}
\caption{Figures a)-d) represent the renormalization of the electromagentic current in $L^{d \leq 4}_{\rm LEFT}$. The later two diagrams illustrate
a penguin diagram c) leading to a surface counter-term in d).}\label{figure1}
\end{figure}

Standard arguments advanced to establish the non-renormalization of $J^\nu $ are concerned with Fig.~\ref{figure1} a)-b). Fig.~\ref{figure1} a) represents wavefunction renormalization, while the insertion of the current
is represented with a circled cross in Fig.~\ref{figure1} b)-c). The divergence and finite terms of diagrams a)-b) cancel at zero momentum transfer for an on-shell state.
For a one electron state, the Noether current corresponds to the electron number current, which we label as $J_N^\nu$ consistent with Ref.~\cite{Collins:2005nj}.
The usual textbook argument then concludes
\bea
\mu \frac{d}{d \mu} J_N^\nu = 0,
\eea
consistent with the current being conserved.
However, the penguin diagram in Fig.~\ref{figure1} c) is divergent. This divergence is cancelled by a counter-term of the form $\partial^\nu F_{\nu \mu}$ shown in Fig.~\ref{figure1} d).
This operator has a four divergence that identically vanishes (i.e. corresponds to a surface term).
The EOM of the $A^{\mu}$ field is given by
\begin{align}
0 =& \frac{\delta S_{\rm LEFT}}{\delta A_{\mu}(x)} = e\mu^\epsilon J_{N}^\mu
+ Z_3 \partial_{\nu}F^{\nu\mu} + \frac{1}{\xi}\partial^\mu \partial\cdot A.
\end{align}
The EOM relates terms in a non-intuitive fashion when an extremum of the action is taken.
$J_{N}^\mu$ receives a multiplicative renormalization generated from the nonzero anomalous dimension of the second term as a result. The current can be subsequently
redefined to remove this effect and cancel the running, as shown in Ref.~\cite{Collins:2005nj}.
\begin{figure}[!ht]
\includegraphics[width=0.175\textwidth]{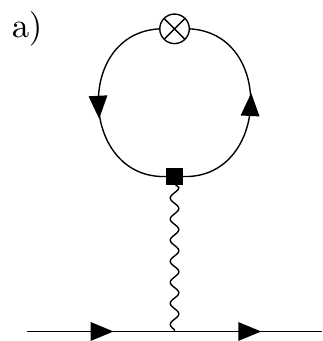}
\includegraphics[width=0.175\textwidth]{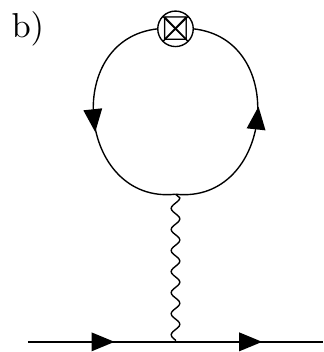}
\caption{Figure a) shows the insertion of a dipole operator in a one loop diagram  (black square) with the $d \leq 4$ LEFT electromagnetic current as a circled cross.
Figure b) shows the insertion of a dipole contribution to the current as a circled cross box.}\label{figure2}
\end{figure}

\begin{figure}[!ht]
\includegraphics[width=0.2\textwidth]{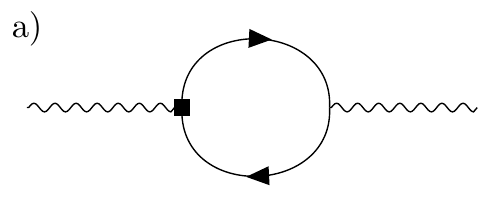}
\includegraphics[width=0.2\textwidth]{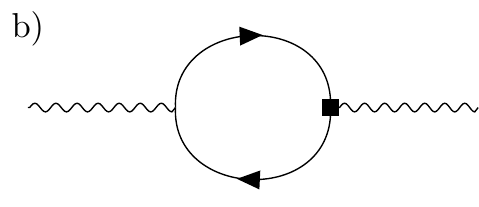}
\linebreak
\includegraphics[width=0.2\textwidth]{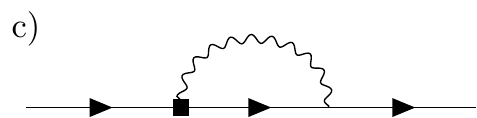}
\includegraphics[width=0.2\textwidth]{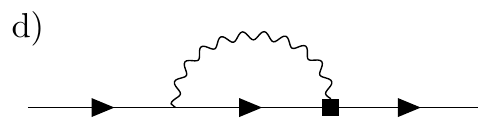}
\linebreak
\includegraphics[width=0.2\textwidth]{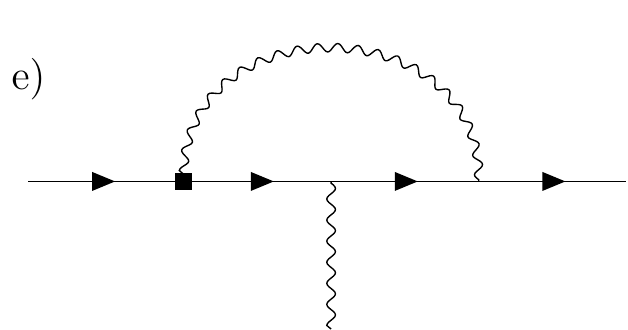}
\includegraphics[width=0.2\textwidth]{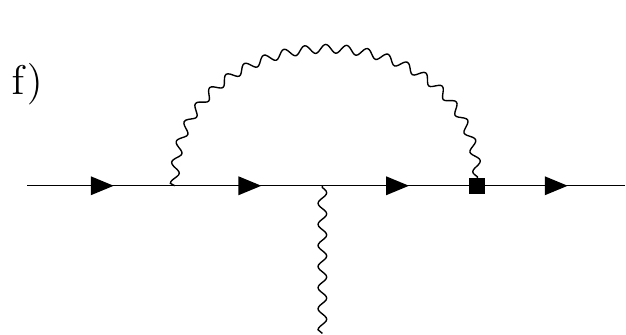}
\linebreak
\includegraphics[width=0.2\textwidth]{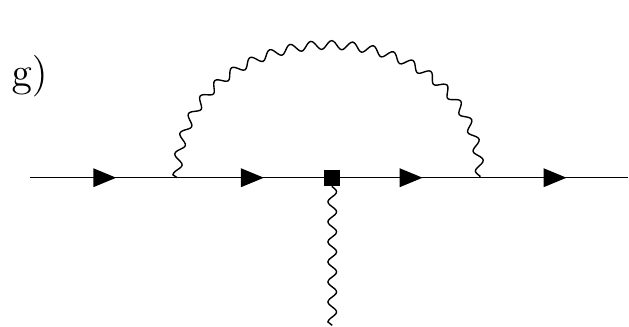}
\caption{One loop diagrams generating the divergences of the LEFT that are removed with the renormalization reported in Ref.~\cite{Jenkins:2017dyc}.}\label{figure3}
\end{figure}

Fig.~\ref{figure2} shows the need to further refine this argument in the presence of higher dimensional operators. 
These diagrams are the direct analogy
to the arguments of Ref.~\cite{Collins:2005nj} leading to a redefinition of the current due to the mixing of the dipole operator 
with the counter-term multiplying $\partial_{\nu}F^{\nu\mu}$.
Inserting the dipole operator (indicated with a black box) with the electromagnetic current, indicated with a circled cross in Fig.~\ref{figure2}a), gives mixing proportional to $M_e/v_T$. 
Including the effect of the dipole operator in the current insertion is indicated by a ``circled cross box" in Fig.~\ref{figure2}b).
Calculating the diagrams directly for an electron in the loop
gives a contribution to the photon two point function of the form
\begin{align}
- \Delta Z_3 &= - \frac{e q_e}{2 \, \pi^2 \, \epsilon} (C_{\substack{e \gamma \\ 11}} [M_e]_{11} + C_{\substack{e \gamma \\ 11}}^\star [M^\dagger_e]_{11}).
\end{align}
This divergence is cancelled by a counter-term \cite{Jenkins:2017dyc} which leads to a modification of $Z_3$ of the form 
$\Delta Z_3$. (The generalization to other charged leptons in the loop is trivial.) This is as expected as a corresponding divergence is present in the LEFT in Fig.~\ref{figure3} a)-b) and the external photon does not play a role that distinguishes
the divergence obtained once the current is redefined.
We have calculated the diagrams in Fig.~\ref{figure3} and agree with the corresponding dipole operator results in Ref.~\cite{Jenkins:2017dyc}.

The interpretation of this mixing effect is subtle in the LEFT.
Varying $S_{\rm LEFT}$ with respect to $A_{\mu}^{(r)}$ gives
\bea
0 &=& \frac{\delta S}{\delta A_{\mu}} = e\mu^\epsilon J_{N}^\mu
+ Z_3 \partial_{\nu}F^{\nu\mu} + \frac{1}{\xi}\partial^\mu \partial\cdot A  \\
&+&  \sqrt{Z_3} Z_2 \partial_\nu \left( Z_C C_{e\gamma}
\left(\overline e_L \sigma^{\nu\mu}e_R\right)
+ Z_C^\star C^*_{e\gamma}\left(\overline e_R \sigma^{\nu\mu} e_L\right)  \right) + \dots \nonumber
\eea
The tree level contributions to the electron number operator of terms $\propto C_{e\gamma},C^*_{e\gamma}$ vanish 
at infinity by Stokes' theorem.\footnote{We thank Mark Wise for discussions on this point.}
We define a $\overline{\rm MS}$-renormalized current
\bea
&&J_{\overline{\rm MS}}^\mu = J_N^\mu + \frac{Z_3 - 1}{e\mu^\epsilon} \partial_\nu F^{\nu\mu}
\\
&&+ \frac{\sqrt{Z_3} Z_2}{e\mu^\epsilon} \partial_\nu \left(Z_C C_{e\gamma}
\left(\overline e_L \sigma^{\nu\mu}e_R\right)
+Z_C^* C^*_{e\gamma}\left(\overline e_R \sigma^{\nu\mu} e_L\right)  \right) + \dots \nonumber
\eea
The $\overline{\rm MS}$-renormalized current expressed in terms of bare quantities is
\bea
	&&J_{\overline{\rm MS}}^\mu = \overline \psi^{(0)} \gamma^\mu \psi^{(0)}
	+ \frac{1-Z_3^{-1}}{e_0} \partial_\nu F^{(0),\nu\mu} 
	\\
	&&+ \frac{1}{e^{(0)}}  \partial_\nu \left(C^{(0)}_{e\gamma}
\left(\overline e^{(0)}_L \sigma^{\nu\mu}e^{(0)}_R\right)
+ C^{*,(0)}_{e\gamma}\left(\overline e^{(0)}_R \sigma^{\nu\mu} e^{(0)}_L\right)  \right) + \dots \nonumber
\eea
The renormalization group flow of the current is
\bea\label{redefinitioncurrent}
\mu\frac{d}{d\mu}J_{\overline{\rm MS}}^{\mu} = 2\gamma_A \frac{1}{e_0 Z_3}\partial_\nu F^{(0),\nu\mu}.
\eea
The $\overline{\rm MS}$-renormalized current depends on the renormalization scale $\mu$ as in the SM case.
The LEFT dipole corrections to the current fall off at infinity when considering the electron number operator. They also vanish from Eqn.~(\ref{redefinitioncurrent}) as separate terms,
which is consistent with this fact. The dipole operators mix into $\partial^\nu F_{\mu \nu}$ proportional
to $M_e/\bar{v}_T$, a correction with a natural interpretation of an electron dipole charge distribution in the LEFT.
In order to extract a conserved electron number which is independent of the renormalization scale,
we redefine the current, including the effect of dipole operators in direct analogy to Ref.~\cite{Collins:2005nj}. 
We define
\bea
	J_{\rm LEFT phys}^\mu &=& J_{\overline{\rm MS}}^\mu - \frac{\Pi(0)}{e\mu^\epsilon}\partial_\nu F^{\nu\mu}, 
\eea
where $\Pi(0)$ is the electron vacuum polarization in the LEFT, including the effects of operators of mass dimension greater than four.
The electron vacuum polarization is still defined in the standard manner, and the current is modified by a redefinition at $q^2 =0$.
\pagebreak
It follows that
\bea
F_{\rm LEFT,phys}^{\nu\mu} &=& \left[1 + \Pi(0) \right]^{1/2} F^{\nu\mu},  \\
e_{\rm LEFT,phys} &=& \left[1 + \Pi(0) \right]^{-1/2} e \mu^{\epsilon}.
\eea
In the $\rm \overline{MS}$ scheme
\bea
\Pi(0) &=& - \frac{e^2}{12 \pi^2} \log \frac{m_e^2}{\mu^2} \\ 
 &+& \frac{e \, q_e}{2 \, \pi^2} (C_{\substack{e \gamma \\ 11}} [M_e]_{11} + C_{\substack{e \gamma \\ 11}}^\star [M^\dagger_e]_{11})\log \frac{m_e^2}{\mu^2} + \dots \nonumber
\eea

From these results one directly defines the time component of the physical current as
\bea
j_{\rm LEFT phys}^0 = \frac{\nabla \cdot {\bf{E}}_{\rm LEFT,phys}}{-e_{\rm LEFT,phys}},
\eea
which is the appropriate generalization of the source in Gauss's law into the LEFT. 
This is a numerically small effect, as the electromagentic dipole operator is constrained~\cite{Dekens:2018pbu}.

To summarize, higher dimensional operators in the LEFT act to change the relationship between the Lagrangian parameter
$e$ and experimental measurements in a manner that corresponds to dipole operators being present in the LEFT. 
This occurs through a modified source term in Gauss's law that reflects the presence of a multipole expansion in the
EFT. 
The tree level dipole contributions to the electron number operator vanish at infinity by Stokes' theorem, but quantum effects necessitates a redefinition 
of the current.

\section{Conclusions}
We have reported the equations of motion for the LEFT including corrections due to dimension six operators. These results are listed in the Appendix.
These corrections lead directly to questions on the meaning of conserved currents in the LEFT. We have examined how the conserved currents of the LEFT
encode symmetry constraints that are manifest or non-linearly realized. 
We have also generalized and embedded the source in Gauss's law into the LEFT, incorporating the effects
of electrically charged particles having dipole operator sources.

\section{acknowledgments}
The authors acknowledge support from the Villum Fonden and the Danish National Research Foundation (DNRF91)
  through the Discovery center. MT thanks Caltech, San Diego and Perimeter Institute for hospitality when part of this work was
  completed. We thank Ilaria Brivio,  Andy Jackson, Aneesh Manohar, Peter Stoffer,  Anagha Vasudevan and Mark Wise for helpful discussions.

\onecolumngrid

\appendix
\newpage
\section{Appendix}
Our operator label notation for the LEFT is largely consistent with Refs.~\cite{Jenkins:2017jig,Jenkins:2017dyc}. We use a different sign convention
on the charge conjugation operator, here $C = - i \gamma^2 \, \gamma^0$, where as in  Refs.~\cite{Jenkins:2017jig,Jenkins:2017dyc}
$C$ is defined with opposite sign. We further introduce the current notation
\bea
S_{\substack{\psi_1\psi_2,L/R \\ st}} &=& \left(\overline{\psi}_{\substack{1,L/R \\ s}} \psi_{\substack{2,R/L \\t}}\right),  \quad
S_{\substack{\psi_1\psi_2,L/R \\ st}}^{A} = \left(\overline{\psi}_{\substack{1,L/R \\ s}}T^A \psi_{\substack{2,R/L \\t}}\right),  \quad
\cS_{\substack{\psi_1\psi_2,L/R \\ st}}^{a,b} = \left(\overline{\psi}^a_{\substack{1,L/R \\ s}} \psi^b_{\substack{2,R/L \\t}}\right),  \quad
\\
J_{\substack{\psi_1\psi_2,L/R \\ st}}^{\alpha} &=& \left(\overline{\psi}_{\substack{1,L/R \\ s}}\gamma^{\alpha} \psi_{\substack{2,L/R \\t}}\right),  \quad
J_{\substack{\psi_1\psi_2,L/R \\ st}}^{\alpha,A} = \left(\overline{\psi}_{\substack{1,L/R \\ s}}\gamma^{\alpha}T^A \psi_{\substack{2,L/R \\t}}\right),  \quad
\\
\cT_{\substack{\psi_1\psi_2,L/R \\ st}}^{\alpha\beta} &=& \left(\overline{\psi}_{\substack{1,L/R \\ s}}\sigma^{\alpha\beta} \psi_{\substack{2,R/L \\t}}\right),  \quad
\cT_{\substack{\psi_1\psi_2,L/R \\ st}}^{\alpha\beta,A} = \left(\overline{\psi}_{\substack{1,L/R \\ s}}\sigma^{\alpha\beta}T^A \psi_{\substack{2,R/L \\t}}\right),  \quad
\eea
where $J_{\substack{\psi\psi,R \\ st}}^{\alpha} \equiv J_{\substack{\psi,R \\ st}}^{\alpha}$
etc.
We also define the currents where one of the fields is charge conjugated

\bea
\tilde S_{\substack{\psi_1\psi_2,L/R \\ st}} &=&  \left(\overline{\psi}_{\substack{1,L/R \\ s}} \psi_{\substack{2,L/R \\t}}\right), \quad
\tilde \cS^{a,b}_{\substack{\psi_1\psi_2,L/R \\ st}} =  \left(\overline{\psi}^a_{\substack{1,L/R \\ s}} \psi^b_{\substack{2,L/R \\t}}\right), \quad
\nonumber \\
\tilde J_{\substack{\psi_1\psi_2,L/R \\ st}}^{\alpha} &=& \left(\overline{\psi}_{\substack{1,L/R \\ s}}\gamma^{\alpha} \psi_{\substack{2,R/L \\t}}\right),  \quad
\tilde{\mathcal{T}}_{\substack{\psi_1\psi_2,L/R \\ st}}^{\alpha\beta} = \left(\overline{\psi}_{\substack{1,L/R\\s}}\sigma^{\alpha\beta} \psi_{\substack{2,L/R \\ t}}\right), \quad
\eea

and similarly for $\tilde J^{\alpha,A}_{\substack{\psi_1\psi_2,R \\ st}}$ etc.

Using these notational conventions, the EOM for the gauge fields from $L^{(5,6)}$  are

\begin{align}
	%
	%
	%
	%
	%
	%
	\frac{\Delta^{\mu,(5)}_{F}}{2} =&
	\sum_{\psi\neq \nu} C_{\substack{ \psi \gamma \\ pr}} \partial_{\nu}\cT_{\substack{\psi,L \\ pr}}^{\nu\mu}
	+ C_{\substack{ \nu \gamma \\ pr}} \partial_{\nu}\tilde \cT_{\substack{\nu^c\nu,L \\ pr}}^{\nu\mu} + \textrm{h.c.}, \\
	%
	%
	%
	%
	%
	%
	\frac{\Delta^{A\mu,(5)}_{G}}{2} =&  \sum \underset{pr}{C_{\psi G}}\left[D_{\nu}, \overline{\psi}_{\substack{L \\ p}}
	\sigma^{\nu\mu} T \, \psi_{\substack{R \\ r}}  \right]^A + \textrm{h.c.}, \\
	%
	%
	%
	%
	%
	%
	\frac{\Delta^{A\mu,(6)}_{G}}{2} =& 3 C_G f^{ABC}\left[ \partial^{\alpha}\left(G_B^{\mu\beta}G_{C\beta\alpha}\right) + g f_{DEC}G_{\alpha\beta}^D G^{E\beta\mu} G_B^{\alpha}\right]
	+C_{\tilde G} f^{ABC}\left[ \partial^{\alpha}\left(G_C^{\mu\beta}\tilde G_{B\beta\alpha}\right) + g f_{DEB}\tilde G_{\beta\alpha}^D G^{E\mu\beta} G_C^{\alpha}\right] \nonumber \\
	& \hspace{-0.75cm}+C_{\tilde G} f^{ABC}\left[ \partial^{\alpha}\left(\tilde G_B^{\mu\beta}G_{C\alpha\beta}\right) + g f_{DEB}G_{\beta\alpha}^D \tilde G^{E\mu\beta} G_C^{\alpha}\right]
	+ \frac{C_{\tilde G}}{2} f^{ABC}\epsilon_{\alpha\beta}^{ \hspace{0.275cm} \gamma\mu}\left[ \partial_{\gamma}\left(\tilde G_B^{\alpha\delta}G_{C\delta\beta}\right) + g f_{DEB}\tilde G_{\delta\gamma}^E G^{D\alpha\delta} G_C^{\beta}\right].
\end{align}
The $\Delta L, \Delta B = 0$, contributions to the EOM from $L^{(5,6)}$ are as follows

\allowdisplaybreaks

\begin{align}
	\Delta^{(5,B,L)}_{e_R,p} =& C^*_{\substack{e\gamma \\ rp}} \sigma^{\alpha\beta}e_{\substack{L \\ r}}F_{\alpha\beta}, \\
	\Delta^{(5,B,L)}_{u_R,p} =& C^*_{\substack{u\gamma \\ rp}} \sigma^{\alpha\beta}u_{\substack{L \\ r}}F_{\alpha\beta}
	+ C^*_{\substack{uG \\ rp}} \sigma^{\alpha\beta}T^Au_{\substack{L \\ r}}G_{\alpha\beta}^A, \\
	\Delta^{(5,B,L)}_{d_R,p} =& C^*_{\substack{d\gamma \\ rp}} \sigma^{\alpha\beta}d_{\substack{L \\ r}}F_{\alpha\beta}
	+ \underset{rp}{C^*_{dG}} \sigma^{\alpha\beta}T^Ad_{\substack{L \\ r}}G_{\alpha\beta}^A, \\
	\Delta^{(5,B,L)}_{\nu_L,p} =& 0,  \\
	\Delta^{(5,B,L)}_{e_L,p} =& C_{\substack{e\gamma \\ pr}} \sigma^{\alpha\beta}e_{\substack{R \\ r}}F_{\alpha\beta}, \\
	\Delta^{(5,B,L)}_{u_L,p} =& C_{\substack{u\gamma \\ pr}} \sigma^{\alpha\beta}u_{\substack{R \\ r}}F_{\alpha\beta}
	+ C_{\substack{uG \\ pr}} \sigma^{\alpha\beta}T^Au_{\substack{R \\ r}}G_{\alpha\beta}^A, \\
	\Delta^{(5,B,L)}_{d_L,p} =& C_{\substack{d\gamma \\ pr}} \sigma^{\alpha\beta}d_{\substack{R \\ r}}F_{\alpha\beta}
	+ C_{ \substack{dG \\ pr}} \sigma^{\alpha\beta}T^Ad_{\substack{R \\ r}}G_{\alpha\beta}^A.
\end{align}

\begin{align}
	%
	%
	%
	%
	%
	%
	%
	%
	%
	%
	%
	\Delta^{(6,B,L)}_{e_R,p} =& \gamma_{\alpha}e_{\substack{R \\r}} \left(2 \, C^{V,RR}_{\substack{e e \\ prst}} \, J_{\substack{e,R \\ st}}^{\alpha}
	+ C^{V,RR}_{\substack{e u \\ prst}} J_{\substack{u,R \\ st}}^{\alpha}
	+ C^{V,RR}_{\substack{e d \\ prst}} J_{\substack{d,R \\ st}}^{\alpha}
	+ \sum_{\psi,\nu} C^{V,LR}_{\substack{\psi e \\ stpr}} J_{\substack{\psi,L \\ st}}^{\alpha} \right)\nonumber \\
	&+ e_{\substack{L \\r}} \left(2 \, C^{S,RR*}_{\substack{e e \\ rpts}}S_{\substack{e,R \\ st}}
	+ C^{S,RR*}_{\substack{e u \\ rpts}}S_{\substack{u,R \\ st}} + C^{S,RR*}_{\substack{e d \\ rpts}} S_{\substack{d,R \\ st}}
  + C^{S,RL*}_{\substack{e u \\ rpts}}S_{\substack{u,L \\ st}}
	+ C^{S,RL*}_{\substack{e d \\ rpts}}S_{\substack{d,L \\ st}}
	\right)
	\nonumber \\
	&+ \sigma_{\alpha\beta}e_{\substack{L \\ r}} \left( C^{T,RR*}_{\substack{e u \\ rpts}}\cT_{\substack{u, R \\ s t}}^{\alpha\beta} + C^{T,RR*}_{\substack{e d \\ rpts}}\cT_{\substack{d, R \\ st}}^{\alpha\beta} \right)
	+\nu_{\substack{L \\ r}} \left(
	C^{S,RL*}_{\substack{\nu edu \\ rpts}} S_{\substack{ud,L \\ st}}
	+ C^{S,RR*}_{\substack{\nu edu \\ rpts}} S_{\substack{ud,R \\ st}}
	\right)
	+ C^{T,RR*}_{\substack{\nu edu \\ rpts}}\sigma^{\alpha\beta}\nu_{\substack{L \\r}}\cT_{\substack{ud,R \\ st}}^{\alpha\beta}
	,   \\
	%
	%
	%
	%
	%
	%
	%
	%
	%
	%
	%
	%
	\Delta^{(6,B,L)}_{u_R,p} =&
	 \gamma_{\alpha}u_{\substack{R \\r}}\left(
	2C^{V,RR}_{\substack{uu \\ prst}} J_{\substack{u,R \\ st}}^{\alpha}
	+ C^{V,RR}_{\substack{eu \\ stpr}} J_{\substack{e,R \\ st}}^{\alpha}
	+ C^{V1,RR}_{\substack{ud \\ prst}} J_{\substack{d,R \\ st}}^{\alpha}
	+ C^{V,LR}_{\substack{\nu u \\ stpr}} J_{\substack{\nu,L \\ st}}^{\alpha}
	+ C^{V,LR}_{\substack{eu \\ stpr}} J_{\substack{e,L \\ st}}^{\alpha}
	+ C^{V1,LR}_{\substack{du \\ stpr}} J_{\substack{d,L \\ st}}^{\alpha}
	+ C^{V1,LR}_{\substack{uu \\ stpr}} J_{\substack{u,L \\ st}}^{\alpha}
	 \right)
	\nonumber \\ &
	+ \gamma_{\alpha}T^Au_{\substack{R \\r}} \left(
	C^{V8,RR}_{\substack{ud \\ prst}} J_{\substack{d,R \\ st}}^{\alpha,A}
	+ C^{V8,LR}_{\substack{uu \\ stpr}} J_{\substack{u,L \\ st}}^{\alpha,A}
	+ C^{V8,LR}_{\substack{du \\ stpr}} J_{\substack{d,L \\ st}}^{\alpha,A}
	\right)
	+ T^Au_{\substack{L \\r}} \left(
	2C^{S8,RR*}_{\substack{uu \\ rpts}} S_{\substack{u,R \\ st}}^{A}
	+ C^{S8,RR*}_{\substack{ud \\ rpts}} S_{\substack{d,R \\ st}}^{A}
	\right)
	\nonumber \\ &
	+ u_{\substack{L \\r}}\left(
	C^{S,RR*}_{\substack{eu \\ tsrp}} S_{\substack{e,R \\ st}}
	+ 2C^{S1,RR*}_{\substack{uu \\ rpts}}  S_{\substack{u,R \\ st}}
	+ C^{S1,RR*}_{\substack{ud \\ rpts}} S_{\substack{d,R \\ st}}
	+ C^{S,RL}_{\substack{eu\\ stpr}} S_{\substack{e,L \\ st}}
	\right)
	+ C^{T,RR*}_{\substack{eu \\ tsrp}}\sigma_{\alpha\beta}u_{\substack{L \\r}}\cT_{\substack{e,R \\ st}}^{\alpha\beta}
	\nonumber \\ &
	  + \gamma_{\alpha}d_{\substack{R \\r}}\left(
	 C^{V,LR*}_{\substack{\nu edu \\ tsrp}} J_{\substack{e\nu,L \\ st}}^{\alpha}
	 + C^{V1,LR*}_{\substack{uddu \\ tsrp}} J_{\substack{du,L \\ st}}^{\alpha}
	\right)
	+ C^{V8,LR*}_{\substack{uddu \\ tsrp}}\gamma^{\alpha}T^Ad_{\substack{R \\r}}J_{\substack{du,L \\ st}}^{\alpha,A}
	\nonumber \\ &
	 + d_{\substack{L \\r}}\left(
	 C^{S,RR*}_{\substack{\nu edu \\ tsrp}}S_{\substack{e\nu,R \\ st}}
	 + C^{S1,RR*}_{\substack{uddu \\ tsrp}}S_{\substack{du,R \\ st}}
	\right)
	+ C^{S8,RR*}_{\substack{uddu \\ tsrp}}T^Ad_{\substack{L \\r}}S_{\substack{du,R \\ st}}^{A}
	+ C^{T,RR*}_{\substack{\nu edu \\ tsrp}}\sigma_{\alpha\beta}d_{\substack{L \\r}}\cT_{\substack{e\nu,R \\ st}}^{\alpha\beta}
	, \\
	%
	%
	%
	%
	%
	%
	%
	%
	%
	%
	%
	\Delta^{(6,B,L)}_{d_R,p} =&
	\gamma_{\alpha}d_{\substack{R \\ r}}  \left( 2\underset{prst}{C^{V,RR}_{dd}}J_{\substack{d,R \\ st}}^{\alpha}
	+ \underset{stpr}{C^{V,RR}_{ed}}J_{\substack{e,R \\ st}}^{\alpha}
	+   \underset{stpr}{C^{V1,RR}_{ud}} J_{\substack{u,R \\ st}}^{\alpha}
	+ \underset{stpr}{C^{V,LR}_{\nu d}} J_{\substack{\nu,L \\ st}}^{\alpha}
	+ \underset{stpr}{C^{V,LR}_{e d}} J_{\substack{e,L \\ st}}^{\alpha}
	+  \underset{stpr}{C^{V1,LR}_{ud}} J_{\substack{u,L \\ st}}^{\alpha}
	+ \underset{stpr}{C^{V1,LR}_{dd}} J_{\substack{d,L \\ st}}^{\alpha}
	\right) \nonumber \\
	&+ \gamma_{\alpha}T^Ad_{\substack{R \\ r }} \left( \underset{stpr}{C^{V8,RR}_{ud}} J_{\substack{u,R \\ st}}^{\alpha, A}
	+ \underset{stpr}{C^{V8,LR}_{ud}} J_{\substack{u,L \\ st}}^{\alpha,A}
	+ \underset{stpr}{C^{V8,LR}_{dd}} J_{\substack{d,L \\ st}}^{\alpha,A}
	\right) +
	T^A d_{\substack{L \\ r}}\left(
	\underset{tsrp}{C^{S8,RR*}_{ud}} S_{\substack{u,R \\ st}}^{A}
	+ 2\underset{rpts}{C^{S8,RR*}_{dd}} S_{\substack{d,R \\ st}}^{A}
	\right)
	\nonumber \\ &+
	d_{\substack{L \\ r}} \left( \underset{tsrp}{C^{S,RR*}_{ed}} S_{\substack{e,R \\ st}}
	+ \underset{tsrp}{C^{S1,RR*}_{ud}} S_{\substack{u,R \\ st}}
	+ 2\underset{rpts}{C^{S1,RR*}_{dd}} S_{\substack{d,R \\ st}}
	 + \underset{stpr}{C_{ed}^{S,RL}}S_{\substack{e,L \\ st}}
     \right) +
	\underset{tsrp}{C^{T,RR*}_{ed}}\sigma_{\alpha\beta}d_{\substack{L \\ r}}\cT_{\substack{e,R \\ st}}^{\alpha\beta}
	\nonumber \\ &+
     \gamma_{\alpha} u_{\substack{R \\ r}} \left(
     \underset{stpr}{C^{V,LR}_{\nu edu}}J_{\substack{\nu e,L \\ st}}^{\alpha}
     + \underset{stpr}{C^{V1,LR}_{uddu}} J_{\substack{ud,L \\ st}}^{\alpha}
     \right)  +
	\underset{stpr}{C^{V8,LR}_{uddu}}\gamma_{\alpha}T^Au_{\substack{R \\ r}}J_{\substack{ud,L \\ st}}^{\alpha,A} +
	u_{\substack{L \\ r}}\left(
	\underset{rpts}{C^{S1,RR*}_{uddu}}S_{\substack{ud,R \\ st}}
	+ \underset{stpr}{C_{\nu edu}^{S,RL}} S_{\substack{\nu e,L \\ st}}
	\right)
	\nonumber \\ &+
\underset{rpts}{C^{S8,RR*}_{uddu}}T^Au_{\substack{ L\\ r}}S_{\substack{ud, R \\ st}}^{A}
	,
	\\
	%
	%
	%
	%
	%
	%
	%
	%
	%
	%
	%
	\Delta^{(6,B,L)}_{\nu_L,p} =&
	\gamma_{\alpha}\nu_{\substack{L \\ r}} \left(  2 \underset{prst}{C^{V,LL}_{\nu\nu}}J_{\substack{\nu,L \\ st}}^{\alpha}
	+ \sum_{\psi \neq \nu}	\underset{prst}{C_{\nu\psi}^{V,LL}}J_{\substack{\psi,L \\st}}^{\alpha}
	+ \sum_{\psi \neq \nu}	\underset{prst}{C_{\nu\psi}^{V,LR}}J_{\substack{\psi,R \\st}}^{\alpha}
	\right)
	+ \gamma_{\alpha} e_{\substack{L \\ r}} \left(
	\underset{prst}{C^{V,LL}_{\nu edu}} J_{\substack{du, L \\ st}}^{\alpha}
	+ \underset{prst}{C^{V,LR}_{\nu edu}} J_{\substack{du, R \\ st}}^{\alpha}
	\right)
	\nonumber \\
	&+ e_{\substack{R \\ r}} \left(
	\underset{prst}{C^{S,RR}_{\nu edu}} S_{\substack{du, L \\ st}}
	+ \underset{prst}{C^{S,RL}_{\nu edu}} S_{\substack{du, R \\ st}}
	\right)
	+ \underset{prst}{C^{T,RR}_{\nu edu}}\sigma_{\alpha\beta}e_{\substack{R \\ r}} \cT_{\substack{du,L \\ st}}^{\alpha\beta}
	, \\
	\Delta^{(6,B,L)}_{e_L,p} =&
	\gamma_{\alpha} e_{\substack{L \\ r}}\left( 2 \underset{prst}{C^{V,LL}_{ee}}J_{\substack{e,L \\ st}}^{\alpha}
	+ \underset{stpr}{C^{V,LL}_{\nu e}}J_{\substack{\nu,L \\ st}}^{\alpha}
	+ \underset{prst}{C^{V,LL}_{eu}} J_{\substack{u,L \\ st}}^{\alpha}
	+ \underset{prst}{C^{V,LL}_{ed}} J_{\substack{d,L \\ st}}^{\alpha}
	+ \sum_{\psi}\underset{prst}{C^{V,LR}_{e\psi}} J_{\substack{\psi,R \\ st}}^{\alpha}
	\right) \nonumber \\
	&+ e_{\substack{R \\ r}} \left(
	2\underset{prst}{C^{S,RR}_{ee}} S_{\substack{e,L \\ st}}
	+ \underset{prst}{C^{S,RR}_{eu}} S_{\substack{u,L \\ st}}
	+ \underset{prst}{C^{S,RR}_{ed}} S_{\substack{d,L \\ st}}
	+ \underset{prst}{C^{S,RL}_{eu}} S_{\substack{u,R \\ st}}
	+ \underset{prst}{C^{S,RL}_{ed}} S_{\substack{d,R \\ st}}
	\right) \nonumber \\
	&+ \sigma_{\alpha\beta} e_{\substack{R \\ r}} \left(
	\underset{prst}{C^{T,RR}_{eu}} \cT_{\substack{u,L \\ st}}^{\alpha\beta}
	+ \underset{prst}{C^{T,RR}_{ed}} \cT_{\substack{d,L \\ st}}^{\alpha\beta}
	\right) 
	+ \gamma_{\alpha}\nu_{\substack{L \\ r}} \left(
	\underset{rpts}{C^{V,LL*}_{\nu edu}} J_{\substack{ud,L \\ st}}^{\alpha}
	+ \underset{rpts}{C^{V,LR*}_{\nu edu}} J_{\substack{ud,R \\ st}}^{\alpha}
	\right)
	, \\
	%
	%
	%
	%
	%
	%
	%
	%
	%
	%
	%
	\Delta^{(6,B,L)}_{u_L,p} =&
	\gamma_{\alpha} u_{\substack{L \\ r}} \left(
	2\underset{prst}{C^{V,LL}_{uu}} J_{\substack{u,L \\ st}}^{\alpha}
	+ \underset{stpr}{C^{V,LL}_{\nu u}} J_{\substack{\nu,L \\ st}}^{\alpha}
	+ \underset{stpr}{C^{V,LL}_{eu}} J_{\substack{e,L \\ st}}^{\alpha}
	+ \underset{prst}{C^{V1,LL}_{ud}} J_{\substack{d,L \\ st}}^{\alpha}
	+ \underset{prst}{C^{V,LR}_{ue}} J_{\substack{e,R \\ st}}^{\alpha}
	+ \underset{prst}{C^{V1,LR}_{uu}} J_{\substack{u,R \\ st}}^{\alpha}
	+ \underset{prst}{C^{V1,LR}_{ud}} J_{\substack{d,R \\ st}}^{\alpha}
	\right) \nonumber \\
	&+ \gamma_{\alpha}T^A u_{\substack{L \\ r}} \left(
	\underset{prst}{C^{V8,LL}_{ud}} J_{\substack{d,L \\ st}}^{\alpha,A}
	+ \underset{prst}{C^{V8,LR}_{uu}} J_{\substack{u, R \\ st}}^{\alpha,A}
	+ \underset{prst}{C^{V8,LR}_{ud}} J_{\substack{d, R \\ st}}^{\alpha,A}
	\right)
	+ \underset{stpr}{C^{T,RR}_{eu}}\sigma_{\alpha\beta}u_{\substack{R \\ r}}\cT_{\substack{e,L \\ st}}^{\alpha\beta}
	\nonumber \\
	&+ u_{\substack{R \\ r}}\left(
	\underset{stpr}{C^{S,RR}_{eu}} S_{\substack{e,L \\ st}}
	+ 2\underset{prst}{C^{S1,RR}_{uu}} S_{\substack{u, L \\ st}}
	+ \underset{prst}{C^{S1,RR}_{ud}} S_{\substack{d, L \\ st}}
	+ \underset{tsrp}{C^{S,RL*}_{eu}} S_{\substack{e, R \\ st}}
	\right)
	+ T^A u_{\substack{R \\ r}} \left(
	2\underset{prst}{C^{S8,RR}_{uu}}S_{\substack{u,L \\ st}}^{A}
	+ \underset{prst}{C^{S8,RR}_{ud}}S_{\substack{d,L \\ st}}^{A}
	\right)
	\nonumber \\
	&+ \gamma_{\alpha}d_{\substack{L \\ r}} \left(
	\underset{tsrp}{C^{V,LL*}_{\nu edu}} J_{\substack{e\nu,L \\ st}}^{\alpha}
	+ \underset{prst}{C^{V1,LR}_{uddu}} J_{\substack{du,R \\ st}}^{\alpha}
	\right)
	+ \underset{prst}{C^{V8,LR}_{uddu}}\gamma_{\alpha}T^Ad_{\substack{ L\\ r}}J_{\substack{du,R \\ st}}^{\alpha,A}
	\nonumber \\ &
	+ d_{\substack{R \\ r}} \left(
	\underset{prst}{C^{S1,RR}_{uddu}}S_{\substack{du,L \\ st}}
	+ \underset{tsrp}{C^{S,RL*}_{\nu edu}}S_{\substack{e\nu,R\\ st}}
	\right)
	+ \underset{prst}{C^{S8,RR}_{uddu}}T^Ad_{\substack{R \\ r}}S_{\substack{du,L \\ st}}^A
	, \\
	\Delta^{(6,B,L)}_{d_L,p} =&
	\gamma_{\alpha} d_{\substack{L \\ r}} \left(
	2\underset{prst}{C^{V,LL}_{dd}} J_{\substack{d,L \\ st}}^{\alpha}
	+ \underset{stpr}{C^{V,LL}_{\nu d}} J_{\substack{\nu,L \\ st}}^{\alpha}
	+ \underset{stpr}{C^{V,LL}_{ed}} J_{\substack{e,L \\ st}}^{\alpha}
	+ \underset{stpr}{C^{V1,LL}_{ud}} J_{\substack{u,L \\ st}}^{\alpha}
	+ \underset{prst}{C^{V,LR}_{de}} J_{\substack{e,R \\ st}}^{\alpha}
	+ \underset{prst}{C^{V1,LR}_{du}} J_{\substack{u,R \\ st}}^{\alpha}
	+ \underset{prst}{C^{V1,LR}_{dd}} J_{\substack{d,R \\ st}}^{\alpha}
	\right) \nonumber \\
	&+ \gamma_{\alpha}T^A d_{\substack{L \\ r}} \left(
	\underset{stpr}{C^{V8,LL}_{ud}} J_{\substack{u,L \\ st}}^{\alpha,A}
	+ \underset{prst}{C^{V8,LR}_{du}} J_{\substack{u,R \\ st}}^{\alpha,A}
	+ \underset{prst}{C^{V8,LR}_{dd}} J_{\substack{d,R \\ st}}^{\alpha,A}
	\right)
	+ \underset{stpr}{C^{T,RR}_{ed}}\sigma_{\alpha\beta} d_{\substack{R \\ r}}\cT_{\substack{e,L \\ st}}^{\alpha\beta}
	\nonumber \\
	&+ d_{\substack{R \\ r}} \left(
	\underset{stpr}{C^{S,RR}_{ed}} S_{\substack{e,L \\ st}}
	+ \underset{stpr}{C^{S1,RR}_{ud}} S_{\substack{u,L \\ st}}
	+ 2\underset{prst}{C^{S1,RR}_{dd}} S_{\substack{d,L \\ st}}
	+ \underset{tsrp}{C^{S,RL*}_{ed}} S_{\substack{e,R \\ st}}
	\right)
	+ T^A d_{\substack{R \\ r}} \left(
	\underset{stpr}{C^{S8,RR}_{ud}}S_{\substack{u,L \\ st}}^{A}
	+ 2\underset{prst}{C^{S8,RR}_{dd}}S_{\substack{d,L \\ st}}^{A}
	\right)
	\nonumber \\ &+
	\gamma_{\alpha} u_{\substack{L \\ r}} \left(
	\underset{stpr}{C^{V,LL}_{\nu edu}}J_{\substack{\nu e,L \\ st}}^{\alpha}
	+ \underset{rpts}{C^{V1,LR*}_{uddu}} J_{\substack{ud,R \\ st}}^{\alpha}
	\right)
	+ \underset{rpts}{C^{V8,LR*}_{uddu}}\gamma_{\alpha}T^A u_{\substack{L \\ r}}J_{\substack{ud,R \\ st}}^{\alpha,A}
	\nonumber \\ &+
	u_{\substack{R \\ r}}\left(
	\underset{stpr}{C^{S,RR}_{\nu edu}} S_{\substack{\nu e,L \\ st}}
	+ \underset{stpr}{C^{S1,RR}_{uddu}} S_{\substack{ud,L \\ st}}
	\right)
	+ \underset{stpr}{C^{T,RR}_{\nu edu}}\sigma_{\alpha\beta} u_{\substack{R \\ r}}\cT_{\substack{\nu e,L \\ st}}^{\alpha\beta}
	+ \underset{stpr}{C^{S8,RR}_{uddu}} T^A u_{\substack{R \\ r}}S_{\substack{ud,L \\ st}}
	.
\end{align}

The $\Delta L \neq 0$, $\Delta B = 0$ contributions to the EOM from $L^{(6)}$ are

\begin{align}
	%
	%
	%
	%
	%
	%
	\Delta^{(6,B,\slashed L)}_{e_R,p} =&
	e_{\substack{L \\r}} \left(
	C^{S,LL}_{\substack{\nu e \\ stpr}}   \tilde S_{\substack{\nu^{c} \nu,L \\ st}}
	+ C^{S,LR*}_{\substack{\nu e \\ tsrp}} \tilde S_{\substack{\nu \nu^c,L \\ st}}
	\right)
	+ C^{T,LL}_{\substack{\nu e \\ stpr}} \sigma_{\alpha\beta}e_{\substack{L \\r}} \tilde{\cT}_{\substack{\nu^c\nu,L \\ st}}^{\alpha\beta}
	+ \gamma_{\alpha}\nu^c_{\substack{L \\r}}\left(
	C^{V,RL*}_{\substack{\nu edu \\ rpts}} J_{\substack{ud,L \\ st}}^{\alpha}
	+ C^{V,RR*}_{\substack{\nu edu \\ rpts}} J_{\substack{ud,R \\ st}}^{\alpha}
	\right)
	, \\
	%
	%
	%
	%
	%
	%
	\Delta^{(6,B,\slashed L)}_{u_R,p} =&
	u_{\substack{L \\r}} \left(
	C^{S,LL}_{\substack{\nu u \\ stpr}} \tilde S_{\substack{\nu^c\nu,L \\ st}}
	+ C^{S,LR*}_{\substack{\nu u \\ tsrp}} \tilde S_{\substack{\nu\nu^c,L \\ st}}
	\right)
	+ C^{T,LL}_{\substack{\nu u \\ stpr}} \sigma_{\alpha\beta}u_{\substack{L \\r}} \tilde{\cT}_{\substack{\nu^c\nu,L \\ st}}^{\alpha\beta}
	+ C^{S,LR*}_{\substack{\nu edu \\ tsrp}} d_{\substack{L \\r}}\tilde S_{\substack{e\nu^c,L \\ st}}
	+ C^{V,RR*}_{\substack{\nu edu \\ tsrp}} \gamma_{\alpha}d_{\substack{R \\r}} \tilde J_{\substack{e\nu^c,R \\ st}}^{\alpha}
	, \\
	%
	%
	%
	%
	%
	%
	\Delta^{(6,B,\slashed L)}_{d_R,p} =&
	d_{\substack{L \\r}}\left(
	C^{S,LL}_{\substack{\nu d \\ stpr}} \tilde S_{\substack{\nu^c\nu,L \\ st}}
	+ C^{S,LR*}_{\substack{\nu d \\ tsrp}}  \tilde S_{\substack{\nu\nu^c,L \\ st}}
	\right)
	+ C^{T,LL}_{\substack{\nu d \\ stpr}} \sigma_{\alpha\beta}d_{\substack{L \\r}} \tilde{\cT}_{\substack{\nu^c\nu,L\\ st}}^{\alpha\beta}
	+ C^{S,LL}_{\substack{\nu edu \\ stpr}} u_{\substack{L \\ r}} \tilde S_{\substack{\nu^c e,L \\ st}}
	\nonumber \\ &
	+ C^{T,LL}_{\substack{\nu edu \\ stpr}} \sigma_{\alpha\beta}u_{\substack{L \\r}} \tilde{\cT}_{\substack{\nu^c e,L \\ st}}^{\alpha\beta}
	+ C^{V,RR}_{\substack{\nu edu \\ stpr}} \gamma_{\alpha}u_{\substack{R \\r}} \tilde J_{\substack{\nu^c e,L \\ st}}^{\alpha}
	, \\
	%
	%
	%
	%
	%
	%
	\Delta^{(6,B,\slashed L)}_{\nu_L,p} =&
	\nu^c_{\substack{L \\ r}}\left(
	2\underset{prst}{C^{S,LL*}_{\nu\nu}} \tilde S_{\substack{\nu^c\nu,L \\ st}}^*
	+ 2\underset{rpts}{C^{S,LL*}_{\nu\nu}} \tilde S_{\substack{\nu\nu^c,L \\ st}}
	+ \sum_{\psi}\left(\underset{prst}{C^{S,LL*}_{\nu \psi}} S_{\substack{\psi,R \\ st}}^*
	+ \underset{rpts}{C^{S,LL*}_{\nu \psi}} S_{\substack{\psi,L \\ st}}
	+\underset{prst}{C^{S,LR*}_{\nu \psi}} S_{\substack{\psi,L \\ st}}^*
	+ \underset{rpts}{C^{S,LR*}_{\nu \psi}} S_{\substack{\psi,R \\ st}}\right)\right)
	\nonumber \\ &
	+ \sigma_{\alpha\beta}\nu^c_{\substack{L \\ r}} \left(
	\underset{prst}{C^{T,LL*}_{\nu e}} \cT_{\substack{e,R \\ st}}^{\alpha\beta *}
	+ \underset{rpts}{C^{T,LL*}_{\nu e}} \cT_{\substack{e,L \\ st}}^{\alpha\beta}
	+\underset{prst}{C^{T,LL*}_{\nu u}} \cT_{\substack{u,R \\ st}}^{\alpha\beta*}
	+ \underset{rpts}{C^{T,LL*}_{\nu u}} \cT_{\substack{u,L \\ st}}^{\alpha\beta}
	+\underset{prst}{C^{T,LL*}_{\nu d}} \cT_{\substack{d,R \\ st}}^{\alpha\beta *}
	+ \underset{rpts}{C^{T,LL*}_{\nu d}} \cT_{\substack{d,L \\ st}}^{\alpha\beta}
	\right)
	\nonumber \\ &
	+ e_{\substack{L \\ r}}^c \left(
	\underset{prst}{C^{S,LL*}_{\nu edu}} S_{\substack{du,R \\ st}}^*
	+\underset{prst}{C^{S,LR*}_{\nu edu}} S_{\substack{du,L \\ st}}^*
	\right)
	+\underset{prst}{C^{T,LL*}_{\nu edu}}\sigma_{\alpha\beta}e_{\substack{L \\ r}}^c \cT_{\substack{du,R \\ st}}^{\alpha\beta *}
	+ \gamma_{\alpha} e_{\substack{R \\ r}}^c \left(
	\underset{prst}{C^{V,RL*}_{\nu edu}} J_{\substack{du, L \\ st}}^{\alpha *}
	+\underset{prst}{C^{V,RR*}_{\nu edu}} J_{\substack{du, R \\ st}}^{\alpha *}
	\right)
	, \\
	%
	%
	%
	%
	%
	%
	\Delta^{(6,B,\slashed L)}_{e_L,p} =&
	e_{\substack{R \\ r}}\left(
	\underset{tsrp}{C^{S,LL*}_{\nu e}} \tilde S_{\substack{\nu\nu^c,L \\ st}}
	+ \underset{stpr}{C^{S,LR}_{\nu e}} \tilde S_{\substack{\nu^c\nu, L \\ st}}
	\right)
	+ \underset{tsrp}{C^{T,LL*}_{\nu e}} \sigma_{\alpha\beta}e_{\substack{R \\ r}}\tilde{\cT}_{\substack{\nu\nu^c,L \\ st}}^{\alpha\beta}
	\nonumber \\ &
	+ \nu^c_{\substack{L \\ r}} \left(
	\underset{rpts}{C^{S,LL*}_{\nu edu}} S_{\substack{ud,L \\ st}}
	+ \underset{rpts}{C^{S,LR*}_{\nu edu}} S_{\substack{ud,R \\ st}}
	\right)
	+ \underset{rpts}{C^{T,LL*}_{\nu edu}}\sigma_{\alpha\beta}\nu^c_{\substack{L \\ r}}\cT_{\substack{ud,L \\ st}}^{\alpha\beta}
	, \\
	%
	%
	%
	%
	%
	%
	\Delta^{(6,B,\slashed L)}_{u_L,p} =&
	u_{\substack{R \\ r}} \left(
	\underset{tsrp}{C^{S,LL*}_{\nu u}} \tilde S_{\substack{\nu\nu^c,L \\ st}}
	+ \underset{stpr}{C^{S,LR}_{\nu u}} \tilde S_{\substack{\nu^c\nu,L \\ st}}
	\right)
	+ \underset{tsrp}{C^{T,LL*}_{\nu u}} \sigma_{\alpha\beta}u_{\substack{R \\ r}}\tilde{\cT}_{\substack{\nu\nu^c,L \\ st}}^{\alpha\beta}
	+ \underset{tsrp}{C^{S,LL*}_{\nu edu}} d_{\substack{R \\ r}}\tilde S_{\substack{e\nu^c,L \\ st}}
	\nonumber \\ &
	+ \underset{tsrp}{C^{T,LL*}_{\nu edu}} \sigma_{\alpha\beta}d_{\substack{R \\ r}}\tilde{\cT}_{\substack{e\nu^c,L \\ st}}^{\alpha\beta}
	+ \underset{tsrp}{C^{V,RL*}_{\nu edu}} \gamma_{\alpha}d_{\substack{L \\ r}}\tilde J_{\substack{e\nu^c,R \\ st}}^{\alpha}
	, \\
	%
	%
	%
	%
	%
	%
	\Delta^{(6,B,\slashed L)}_{d_L,p} =&
	d_{\substack{R \\ r}} \left(
	\underset{tsrp}{C^{S,LL*}_{\nu d}} \tilde S_{\substack{\nu\nu^c,L \\ st}}
	+ \underset{stpr}{C^{S,LR}_{\nu d}} \tilde S_{\substack{\nu^c\nu,L \\ st}}
	\right)
	+ \underset{tsrp}{C^{T,LL*}_{\nu d}} \sigma_{\alpha\beta}d_{\substack{R \\ r}}\tilde\cT_{\substack{\nu\nu^c,L \\ st}}^{\alpha\beta}
	+ \underset{stpr}{C^{S,LR}_{\nu edu}} u_{\substack{ R \\ r}}\tilde S_{\substack{\nu^c e,L \\ st}}
	+ \underset{stpr}{C^{V,RL}_{\nu edu}} \gamma_{\alpha}u_{\substack{L \\ r}}\tilde J_{\substack{\nu^c e,L \\ st}}^{\alpha}.
\end{align}

 $\Delta L,\Delta B \neq 0$, contributions to the EOM from $L^{(6)}$ are

\begin{align}
	%
	%
	%
	%
	%
	%
	\Delta^{(6,\slashed B,\slashed L)}_{e_R,p} =&
	C^{S,LR*}_{\substack{uud \\ tsrp}}\epsilon_{\alpha\beta\gamma}d_{\substack{R \\r}}^{\gamma c}\tilde\cS_{\substack{u,L \\ st}}^{\beta,\alpha c}
	+ \epsilon_{\alpha\beta\gamma}u_{\substack{R \\r}}^{\gamma c} \left(
	C^{S,LR*}_{\substack{duu \\ tsrp}}\tilde\cS_{\substack{ud,L \\ st}}^{\beta,\alpha c}
	+ C^{S,RR*}_{\substack{duu \\ tsrp}}\tilde\cS_{\substack{ud,R \\ st}}^{\beta,\alpha c}
	\right)
	+ \epsilon_{\alpha\beta\gamma}d_{\substack{L \\r}}^{\gamma } \left(
	C^{S,LL}_{\substack{ddd \\stpr}}\tilde\cS_{\substack{d,L\\ st}}^{\alpha c, \beta}
	+ C^{S,RL}_{\substack{ddd \\ stpr}}\tilde\cS_{\substack{d,R \\ st}}^{\alpha c, \beta}
	\right)
	, \\
	%
	%
	%
	%
	%
	%
	\Delta^{(6,\slashed B,\slashed L)}_{u_R,p} =&
	\epsilon_{\beta\gamma\alpha} e_{\substack{R \\ r}}^{ c}\left(
	\underset{stpr}{C^{S,LR*}_{duu}}\tilde\cS_{\substack{du,L \\ st}}^{\beta c,\gamma *}
	+ \underset{stpr}{C^{S,RR*}_{duu}}\tilde\cS_{\substack{du,R \\ st}}^{\beta c,\gamma *}
	\right)
	+ \epsilon_{\alpha\beta\gamma} u_{\substack{R \\ r}}^{\beta c}\left(
	\underset{prst}{C^{S,RL*}_{uud}}\tilde\cS_{\substack{de,L \\ st}}^{\gamma c,\,\,*}
	- \underset{rpts}{C^{S,RL*}_{uud}}\tilde\cS_{\substack{ed,L \\ st}}^{\,\, , \gamma c}
	\right)
	\nonumber \\ &
	+ \epsilon_{\beta\alpha\gamma} d_{\substack{R \\ r}}^{\beta c}\left(
	\underset{rpts}{C^{S,RL*}_{duu}} \tilde\cS_{\substack{eu,L \\ st}}^{\,\,,\gamma c}
	+ \underset{rpts}{C^{S,RR*}_{duu}} \tilde\cS_{\substack{eu,R \\ st}}^{\,\,,\gamma c}
	+ \underset{rpts}{C^{S,RL*}_{dud}} \tilde\cS_{\substack{\nu d,L \\ st}}^{\,\,,\gamma c}
	- \underset{prst}{C^{S,RR*}_{udd}} \cS_{\substack{\nu d,L \\ st}}^{\,\,,\gamma *}
	\right)
	+ \underset{tsrp}{C^{S,LR*}_{ddu}}\epsilon_{\beta\gamma\alpha} \nu_{\substack{L \\ r}}\tilde\cS_{\substack{d,L \\ st}}^{\gamma,\beta c} ,
	\\
	%
	%
	%
	%
	%
	%
	\Delta^{(6,\slashed B,\slashed L)}_{d_R,p} =&
	\epsilon_{\alpha\beta\gamma} d_{\substack{R \\ r}}^{\beta c}\left(
	\underset{prst}{C^{S,RL*}_{ddu}} \tilde\cS_{\substack{u\nu,L \\ st}}^{\gamma c,\,\,*}
	- \underset{rpts}{C^{S,RL*}_{ddu}} \tilde\cS_{\substack{\nu u,L \\ st}}^{\,\,,\gamma c}
	+ \underset{prst}{C^{S,RL*}_{ddd}} \cS_{\substack{ed,R \\ st}}^{\,\,,\gamma *}
	- \underset{rpts}{C^{S,RL*}_{ddd}} \cS_{\substack{de,L \\ st}}^{\gamma,\,\,}
	+ \underset{prst}{C^{S,RR*}_{ddd}} \cS_{\substack{ed,L \\ st}}^{\,\,,\gamma *}
	- \underset{rpts}{C^{S,RR*}_{ddd}} \cS_{\substack{de,R \\ st}}^{\gamma,\,\,}
	\right)
	\nonumber \\ &
	+ \epsilon_{\alpha\beta\gamma} u_{\substack{R \\ r}}^{\beta c}\left(
	 \underset{prst}{C^{S,RL*}_{duu}} \tilde\cS_{\substack{ue,L \\ st}}^{\gamma c,\,\,*}
	 + \underset{prst}{C^{S,RR*}_{duu}} \tilde\cS_{\substack{ue,R \\ st}}^{\gamma c,\,\,*}
	 + \underset{prst}{C^{S,RL*}_{dud}} \tilde\cS_{\substack{d\nu,L \\ st}}^{\gamma c,\,\,*}
	 - \underset{rpts}{C^{S,RR*}_{udd}} \cS_{\substack{d\nu,R \\ st}}^{\gamma,\,\,}
	\right)
	+ \underset{stpr}{C^{S,LR*}_{uud}}\epsilon_{\beta\gamma\alpha} e_{\substack{R \\ r}}^{c}\tilde\cS_{\substack{u,L \\ st}}^{\beta c,\gamma *}
	\nonumber \\ &
	+ \epsilon_{\beta\gamma\alpha} e_{\substack{L \\ r}}\left(
	\underset{tsrp}{C^{S,LR*}_{ddd}} \tilde\cS_{\substack{d,L \\ st}}^{\gamma,\beta c}
	+ \underset{tsrp}{C^{S,RR*}_{ddd}} \tilde\cS_{\substack{d,R \\ st}}^{\gamma,\beta c}
	\right)
	+ \epsilon_{\beta\gamma\alpha} \nu_{\substack{L \\ r}}\left(
	\underset{tsrp}{C^{S,LR*}_{udd}} \tilde\cS_{\substack{du,L \\ st}}^{\gamma,\beta c}
	+ \underset{tsrp}{C^{S,RR*}_{udd}} \tilde\cS_{\substack{du,R \\ st}}^{\gamma,\beta c}
	\right)
	, \\
	%
	%
	%
	%
	%
	%
	\Delta^{(6,\slashed B,\slashed L)}_{\nu_L,p} =&
	\epsilon_{\alpha\beta\gamma} d_{\substack{L \\ r}}^{\gamma c}\left(
	\underset{tsrp}{C^{S,LL*}_{udd}} \tilde\cS_{\substack{du,L \\ st}}^{\beta,\alpha c}
	+ \underset{tsrp}{C^{S,RL*}_{dud}} \tilde\cS_{\substack{ud,R \\ st}}^{\beta,\alpha c}
	\right)
	+ \epsilon_{\alpha\beta\gamma} d_{\substack{R \\ r}}^{\gamma } \left(
	\underset{stpr}{C^{S,LR}_{udd}} \tilde\cS_{\substack{ud,L \\ st}}^{\alpha c,\beta}
	+ \underset{stpr}{C^{S,RR}_{udd}} \tilde\cS_{\substack{ud,R \\ st}}^{\alpha c,\beta}
	\right)
	\nonumber \\ &
	+ \underset{tsrp}{C^{S,RL*}_{ddu}}\epsilon_{\alpha\beta\gamma} u_{\substack{L \\ r}}^{\gamma c}\tilde\cS_{\substack{d,R \\ st}}^{\beta,\alpha c}
	+ \underset{stpr}{C^{S,LR}_{ddu}}\epsilon_{\alpha\beta\gamma} u_{\substack{R \\ r}}^{\gamma } \tilde\cS_{\substack{d,L \\ st}}^{\alpha c,\beta}
	, \\
	%
	%
	%
	%
	%
	%
	\Delta^{(6,\slashed B,\slashed L)}_{e_L,p} =&
	\epsilon_{\alpha\beta\gamma} u_{\substack{L \\ r}}^{\gamma c} \left(
	\underset{tsrp}{C^{S,LL*}_{duu}}\tilde\cS_{\substack{ud,L \\ st}}^{\beta,\alpha c}
	+ \underset{tsrp}{C^{S,RL*}_{duu}}\tilde\cS_{\substack{ud,R \\ st}}^{\beta,\alpha c}
	\right)
	+ \underset{tsrp}{C^{S,RL*}_{uud}}\epsilon_{\alpha\beta\gamma} d_{\substack{L \\ r}}^{\gamma c}\tilde\cS_{\substack{u,R \\ st}}^{\beta,\alpha c}
	+ \epsilon_{\alpha\beta\gamma} d_{\substack{R \\ r}}^{\gamma } \left(
	\underset{stpr}{C^{S,LR}_{ddd}}\tilde\cS_{\substack{d,L \\ st}}^{\alpha c,\beta}
	+ \underset{stpr}{C^{S,RR}_{ddd}}\tilde\cS_{\substack{d,R \\ st}}^{\alpha c,\beta}
	\right)
	, \\
	%
	%
	%
	%
	%
	%
	\Delta^{(6,\slashed B,\slashed L)}_{u_L,p} =&
	\epsilon_{\alpha\beta\gamma} d_{\substack{L \\ r}}^{\beta c} \left(
	\underset{prst}{C^{S,LL*}_{udd}} \tilde\cS_{\substack{d\nu,L \\ st}}^{\gamma c, \, \, *}
	+ \underset{prst}{C^{S,LR*}_{udd}} \cS_{\substack{\nu d,L \\ st}}^{\,\,,\gamma *}
	- \underset{rpts}{C^{S,LL*}_{duu}} \tilde\cS_{\substack{eu,L \\ st}}^{\,\,,\gamma c}
	- \underset{rpts}{C^{S,LR*}_{duu}} \tilde\cS_{\substack{eu,R \\ st}}^{\,\,,\gamma c}
	\right)
	+ \underset{stpr}{C^{S,RL}_{ddu}}\epsilon_{\beta\gamma\alpha} \nu_{\substack{L \\ r}}^{c} \tilde\cS_{\substack{d,R \\ st}}^{\beta c,\gamma *}
	\nonumber \\ &
	+ \epsilon_{\beta\alpha\gamma} e_{\substack{L \\ r}}^{c} \left(
	\underset{stpr}{C^{S,LL*}_{duu}} \tilde\cS_{\substack{du,L \\ st}}^{\beta c,\gamma *}
	+ \underset{stpr}{C^{S,RL*}_{duu}} \tilde\cS_{\substack{du,R \\ st}}^{\beta c,\gamma *}
	\right)
	+ \epsilon_{\alpha\beta\gamma} u_{\substack{L \\ r}}^{\beta c}\left(
	\underset{prst}{C^{S,LR*}_{uud}} \tilde\cS_{\substack{de,R \\ st}}^{\gamma c, \,\, *}
	- \underset{rpts}{C^{S,LR*}_{uud}} \tilde\cS_{\substack{ed,R \\ st}}^{\,\,,\gamma c}
	\right)
	, \\
	%
	%
	%
	%
	%
	%
	\Delta^{(6,\slashed B,\slashed L)}_{d_L,p} =&
	\epsilon_{\beta\alpha\gamma} u_{\substack{L \\ r}}^{\beta c}\left(
	\underset{rpts}{C^{S,LL*}_{udd}} \tilde\cS_{\substack{\nu d,L \\ st}}^{\,\,,\gamma c}
	+ \underset{rpts}{C^{S,LR*}_{udd}} \cS_{\substack{d\nu,R \\ st}}^{\gamma,\,\,}
	- \underset{prst}{C^{S,LL*}_{duu}} \tilde\cS_{\substack{ue,L \\ st}}^{\gamma c,\,\, *}
	- \underset{prst}{C^{S,LR*}_{duu}} \tilde\cS_{\substack{ue,R \\ st}}^{\gamma c,\,\, *}
	\right)
	+ \underset{stpr}{C^{S,RL*}_{uud}}\epsilon_{\beta\gamma\alpha} e_{\substack{L \\ r}}^{c} \tilde\cS_{\substack{u,R \\ st}}^{\beta c,\gamma *}
	\nonumber \\ &
	+ \epsilon_{\beta\gamma\alpha} e_{\substack{R \\ r}}\left(
	\underset{tsrp}{C^{S,LL*}_{ddd}} \tilde\cS_{\substack{d,L \\ st}}^{\gamma,\beta c}
	+ \underset{tsrp}{C^{S,RL*}_{ddd}} \tilde\cS_{\substack{d,R \\ st}}^{\gamma,\beta c}
	\right)
	+ \epsilon_{\beta\gamma\alpha} \nu_{\substack{L \\ r}}^{c}\left(
	\underset{stpr}{C^{S,LL*}_{udd}} \tilde\cS_{\substack{ud,L \\ st}}^{\beta c,\gamma *}
	+ \underset{stpr}{C^{S,RL*}_{dud}} \tilde\cS_{\substack{du,R \\ st}}^{\beta c,\gamma *}
	\right)
	\nonumber \\ &
	+ \epsilon_{\alpha\beta\gamma} d_{\substack{L \\ r}}^{\beta c}\left(
	 \underset{prst}{C^{S,LL*}_{ddd}} \cS_{\substack{ed,R \\ st}}^{\,\,,\gamma *}
	 - \underset{rpts}{C^{S,LL*}_{ddd}} \cS_{\substack{de,L \\ st}}^{\gamma,\,\,}
	 + \underset{prst}{C^{S,LR*}_{ddu}} \cS_{\substack{\nu u,L \\ st}}^{\,\,,\gamma *}
	 - \underset{rpts}{C^{S,LR*}_{ddu}} \cS_{\substack{u\nu,R \\ st}}^{\gamma,\,\,}
	 + \underset{prst}{C^{S,LR*}_{ddd}} \cS_{\substack{ed,L \\ st}}^{\,\,,\gamma *}
	 - \underset{rpts}{C^{S,LR*}_{ddd}} \cS_{\substack{de,R \\ st}}^{\gamma,\,\,}
	\right)
	.
\end{align}
Finally, the dimension 3 and 5 LEFT operators contributing to the neutrino EOM give

\begin{align}
	%
	%
	%
	%
	%
	%
	\Delta^{(3)}_{\substack{\nu_L \\ p}} &= - 2\underset{pr}{C^*_{\nu}}\nu^c_{\substack{L \\ r}}, \\
	%
	%
	%
	%
	%
	%
	\Delta^{(5)}_{\substack{\nu_L \\ p}} &= 2 \, C^*_{\substack{\nu\gamma \\ pr}}\sigma^{\alpha\beta}\nu_{\substack{L \\ r}}^c F_{\alpha\beta}.
\end{align}


\newpage
\bibliography{bibliography_V2-1}

\end{document}